\documentclass[aps,pre,twocolumn,twoside,superscriptaddress,floatfix, a4paper,showpacs]{revtex4}
\usepackage{amsmath}
\usepackage{amsfonts}
\usepackage{amssymb}
\usepackage{graphicx,epsfig}

\begin{document}

\title{Super-aging correlation function and ergodicity breaking for Brownian Motion in logarithmic potentials}

\author{A. Dechant}
\affiliation{Department of Physics, University of Augsburg, 86135 Augsburg, Germany}
\affiliation{Dahlem Center for Complex Quantum Physics, FU Berlin, 14195 Berlin, Germany}
\author{E. Lutz}
\affiliation{Department of Physics, University of Augsburg, 86135 Augsburg, Germany}
\affiliation{Dahlem Center for Complex Quantum Physics, FU Berlin, 14195 Berlin, Germany}
\author{D.A. Kessler}
\affiliation{Department of Physics, Bar Ilan University, Ramat-Gan 52900, Israel}
\author{E. Barkai}
\affiliation{Department of Physics, Bar Ilan University, Ramat-Gan 52900, Israel}

\begin{abstract}
We consider an overdamped Brownian particle moving in a confining asymptotically logarithmic potential, which supports a normalized Boltzmann equilibrium density. We derive analytical expressions for  the two-time correlation function and the fluctuations of the time-averaged position of the particle for large but finite times. We characterize the occurrence of aging and nonergodic behavior as a function of the depth of the potential, and we support our predictions with extensive Langevin simulations. While the Boltzmann measure is used to obtain stationary correlation functions, we show how the non-normalizable infinite covariant density is related to the super-aging behavior. \end{abstract}

\maketitle

\section{Introduction} \label{SEC_1}

Ergodicity is a central concept in the theory of stochastic processes. A random variable $A$ is said to be ergodic when its time average $\bar{A}(t) = (1/t) \int_{0}^{t} dt' A(t')$  over a single realization in the limit of infinitely long times is  equal to the equilibrium ensemble average $\langle A \rangle = \int dA \,A P(A)$ over many realizations of the process \cite{Pap91}. Here $P(A)$ is the equilibrium probability density for the random variable $A$. For ergodic variables, the width of the probability distribution of the random time average vanishes in the long-time limit, and the distribution reduces asymptotically to a delta function centered on the ensemble average, $Q \left( \bar{A},t \right) \rightarrow  \delta \left( \bar{A} - \langle A \rangle \right)$. For nonergodic variables, the time average remains a stochastic quantity even in the infinite time limit. A general criterion for the ergodicity of a process is given by the Khinchin theorem \cite{Khi49}, which asserts that a stationary process is ergodic if its autocorrelation function $\langle A(t) A(t_0) \rangle \rightarrow \langle A \rangle^2$ when $ | t-t_0 | \rightarrow \infty$. However, for processes for which a stationary autocorrelation function does not exist, we cannot use the Khinchin theorem to predict the ergodic properties of the process. For specific systems, there are generalizations of the Khinchin theorem to the nonstationary case \cite{Bur10,Wer10}, but for a general system, the ergodic properties are not straightforward to predict. Meanwhile, for finite times, the time average will be a random variable for any process, ergodic or  not. Since in all experiments the measurement time might be large, but is always finite, it is essential to determine the properties of the distribution of the time average, as the latter cannot be evaluated using 
the equilibrium measure $P(A)$. The distribution of the time average has been investigated  for continuous-time random-walk models \cite{Reb07}, but no general theory exists. In Ref.~\cite{Dec11},  we have provided a general expression for the variance of the time-averaged position $\bar{x}(t)$ for an overdamped Brownian particle in a binding field and showed that, in the special case of a logarithmic potential, the ergodic hypothesis breaks down.

In the present paper, we study in more detail the variance $\bar{\sigma}^2(t) = \langle \bar{x}^2(t) \rangle - \langle \bar{x}(t) \rangle^2$ of the time-averaged position of an overdamped  Brownian particle moving in a confining potential that is asymptotically logarithmic. This system defines an  important class of processes which has found widespread applications in the description of the dynamics of particles near a long, charged polymer \cite{Man69}, momentum diffusion in dissipative optical lattices \cite{Cas90,Mar96,Kat97,Lut03,Dou06,Sag11,Bar12,Dec12}, probe particles in one-dimensional driven fluids \cite{Lev05}, self-gravitating Brownian particles \cite{Cha10}, long-range interacting systems \cite{Bou05,Bou05a,Cha07}, and diffusion of fractals \cite{Vla94}, as well as the dynamics of bubbles in DNA molecules \cite{Fog07,Fog07a,Bar07,Wu09}, of vortices \cite{Bra00}, and of trapped nanoparticles  \cite{Coh05}. A striking characteristic of these systems is that the equilibrium probability distribution possesses a power-law tail  which may lead to diverging moments and ergodicity breaking \cite{Lut04}. Their anomalous behavior is controlled by a single parameter  that is related to the depth of the potential. In the following, we derive explicit long-time expressions for the two-time correlation function and for the time-averaged variance of the position of the particle, and we investigate in detail the diffusive and ergodic properties of the system for large, but finite times. We show how a super-aging correlation function describes this system, even when the stationary Boltzmann distribution is normalizable.

\begin{figure}
\includegraphics[trim=25mm 5mm 25mm 10mm, clip, width=0.47\textwidth]{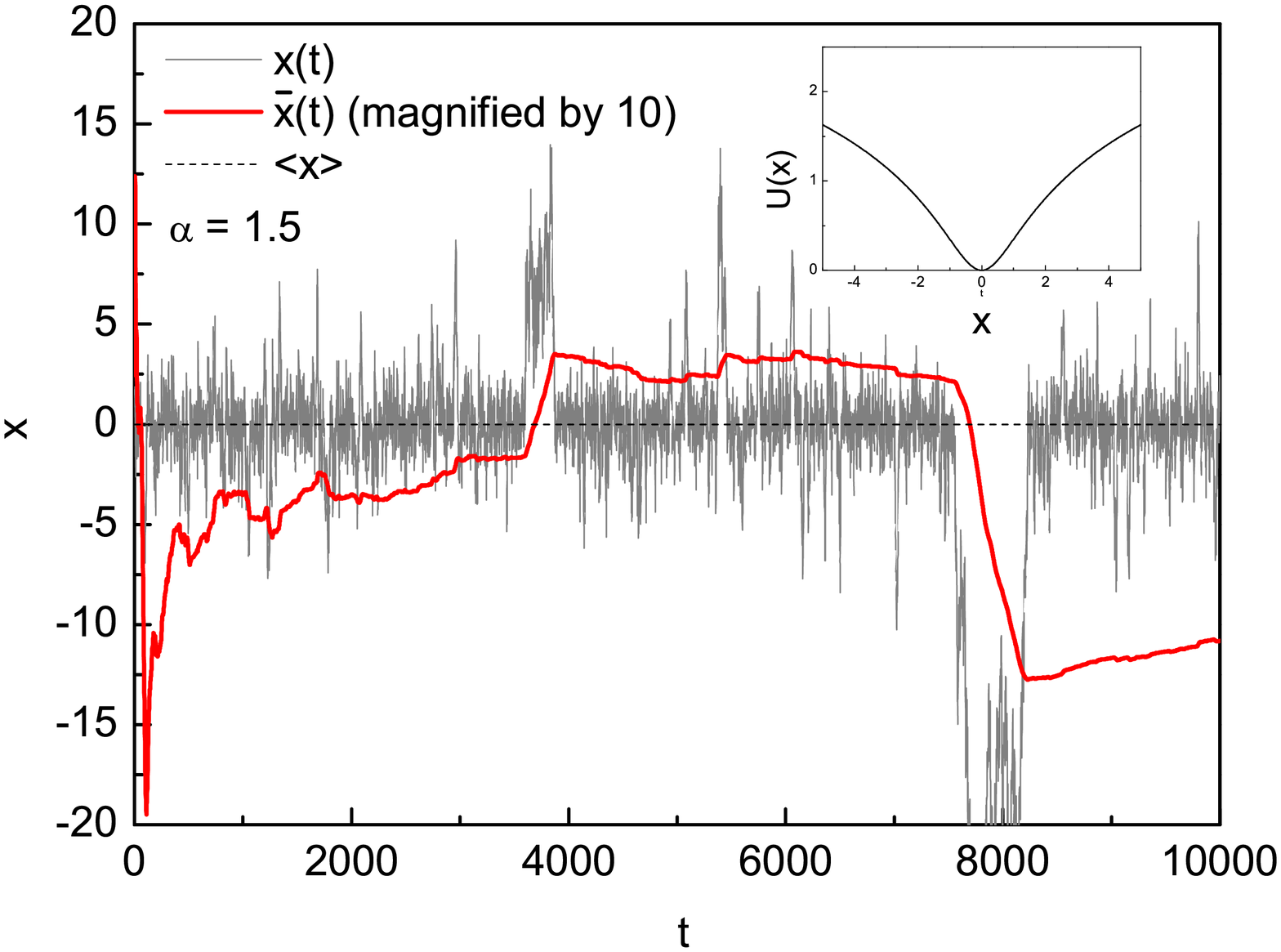}
\includegraphics[trim=25mm 5mm 25mm 10mm, clip, width=0.47\textwidth]{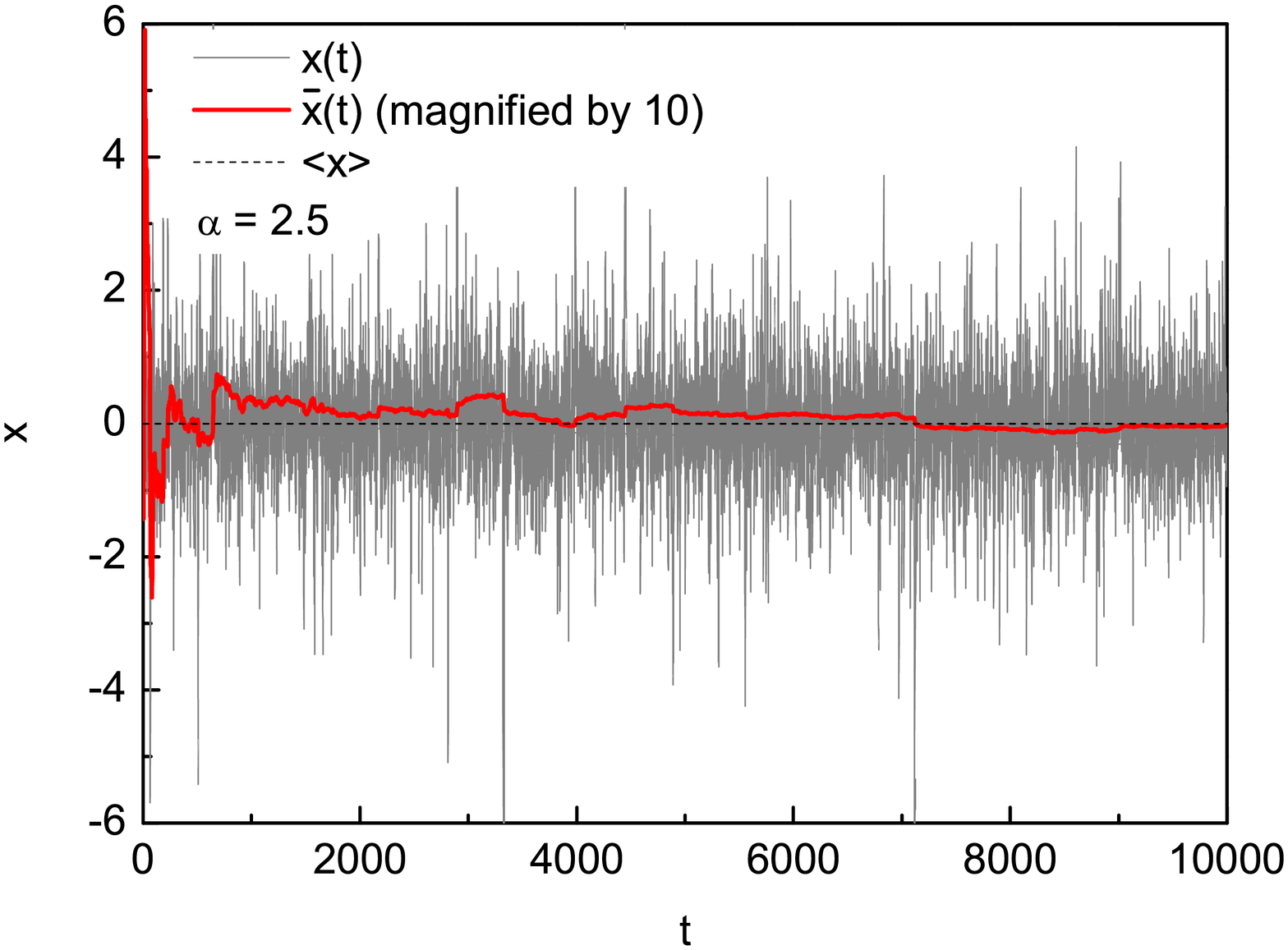}
\caption{(Color online) Typical trajectories (thin gray) and their time average (thick red) for the Brownian particle moving in an asymptotically logarithmic potential $U(x)=(U_0/2) \ln(1+x^2)$ (top panel, inset). The parameter $\alpha$ measures the ratio of the depth of the potential $U_0$ and the temperature [see Eq.~\eqref{A_12}], here we used $k_B T = 0.5$, $\gamma = 1$. The top panel shows the nonergodic phase ($U_0=2 k_B T$), where the time average $\bar{x}(t)$ does not converge to the ensemble average $\langle x \rangle$ (which is zero, dashed line).  The bottom panel shows the ergodic phase ($U_0=4 k_B T$); here the time average does tend to the ensemble average for long times. Also note the different scales for the $x$-axes.}
\label{Fig_1_1}
\end{figure}

Figure \ref{Fig_1_1} shows a typical trajectory of a Brownian particle in a logarithmic potential in the nonergodic phase, obeying the overdamped Langevin equation
\begin{align}
\frac{d x}{d t} = - \frac{1}{m \gamma} \frac{\partial U(x)}{\partial x} + F(t) \label{lang}
\end{align} 
with $U(x) = (U_0/2) \ln(1+x^2)$ and the fluctuating Gaussian noise $F(t)$, $\langle F(t) F(t') \rangle = (2 k_B T/(m \gamma)) \delta(t-t')$. Here, $m$ is the mass of the particle, $\gamma$ is the friction coefficient, $k_B$ is the Boltzmann constant and $T$ is the temperature.
Obviously the time average $\bar{x}(t) = (1/t) \int_{0}^{t} dt' x(t')$ does not converge to the ensemble average $\langle x \rangle=0$ even for long times. The reason for this behavior is the long excursions of the particle into the tails of the potential, where the slope and thus the restoring force tends to zero for $x \gg 1$. On average, these excursions get ever longer as time increases and thus dominate the time average for all times. Our goal is to characterize the nonergodic behavior of this process.

The outline of the paper is as follows. In Section \ref{SEC_2}, we  solve the Fokker-Planck equation corresponding to Eq.~\eqref{lang} for the Brownian particle by transforming it into  a Schr\"odinger-like equation and using an  eigenfunction expansion. In Section \ref{SEC_3}, we explicitly construct the conditional probability density from the eigenfunction expansion and determine its asymptotic long-time behavior, as well as those of the first and second moments of the position variable.  We then use the conditional probability density in Section \ref{SEC_4} to compute the two-time position correlation function, which exhibits either stationary behavior or nonstationary, super-aging behavior, depending on the depth of the potential. Finally, in Section \ref{SEC_5}, we evaluate the variance of the time-averaged position and show the existence of a threshold for the potential depth below which ergodicity is broken. Parts of our results were obtained in Ref.~\cite{Dec11} with the help of a scaling ansatz. We provide here a complementary and more detailed derivation.

\section{Fokker-Planck equation for the logarithmic potential} \label{SEC_2}

We consider an overdamped Brownian particle moving in a symmetric ($U(x)=U(-x)$) potential $U(x)$ that is asymptotically logarithmic for large $|x|\gg a$,  $U(x)\simeq  U_0 \ln(|x|/a)$; during the calculation, we will set $a=1$. For $a \neq 1$, the variable $x$ in our results should be replaced by $x/a$. An example for such a potential is $U(x) = (U_0/2) \ln(1+x^2)$, which is what we use for our numerical Langevin simulations. The dynamics of the particle is governed by the Fokker-Planck equation for the probability density $W(x,t)$ corresponding to the Langevin equation \eqref{lang},
\begin{align}
\frac{\partial}{\partial t} W(x,t) &= \frac{D}{k_B T} \frac{\partial}{\partial x} \left(U'(x) W(x,t) \right) + D \frac{\partial^2}{\partial x^2} W(x,t) \label{A_3a} \ , 
\end{align}
where 
$D=k_BT/(m \gamma)$ is the diffusion coefficient.
In the following, we set the mass of the particle to $m=1$. The stationary equilibrium solution to Eq.~\eqref{A_3a} is given by $W_{\text{eq}}(x) = \exp(-U(x)/(k_B T))/Z$ with the normalizing partition function $Z = \int dx \exp(-U(x)/(k_B T))$. As we will see, the partition function $Z$ and thus the stationary probability density do not always exist for the logarithmic potential. Since for the logarithmic potential, we have asymptotically $U'(x) \simeq U_0/|x|$, both the first (drift) and second (diffusion) term on the right-hand side of Eq.~\eqref{A_3a} scale as $1/x^2$ \cite{Bar10}. This scaling is responsible for the interesting effects discussed in the following. The addition of a linear force to the logarithmic potential breaks this scaling and leads to a very different behavior \cite{Arv06,Bez06,Fur07,Sil11}. 

Our goal is to evaluate  the  long-time behavior of the variance  of the time-averaged position,  $\bar{x}(t) =  \int_{0}^{t} dt' \, x(t') /t$, which is given by,
\begin{align}
\bar{\sigma}^2(t) &= \langle \bar{x}^2(t) \rangle - \langle \bar{x}(t) \rangle^2 \nonumber \\
&= \frac{1}{t^2} \left[ \int_{0}^{t} dt'' \int_{0}^{t} dt' \, C(t'',t') \right. \nonumber \\
& \quad \left. - \left( \int_{0}^{t} dt' \langle x(t') \rangle \right)^2 \right] \label{2} \ .
\end{align}
For an ergodic system, we expect $\bar{x}(t) \rightarrow \langle x \rangle_{\text{eq}} = 0$ and thus $\bar{\sigma}^2(t) \rightarrow 0$.
The two-time position correlation function $C(t,t_0) = \langle x(t) x(t_0) \rangle $ in Eq.~\eqref{2} can be expressed in the form \cite{Ris86},
\begin{align}
C(t,t_0) &= \int_{-\infty}^{\infty} dx \int_{-\infty}^{\infty} dx_0 \, x x_0 P(x,t \vert x_0, t_0) W(x_0, t_0)  \ ,\label{1}
\end{align}
where $P(x,t \vert x_0,t_0)$ is the conditional probability density. 
This expression is correct for a prior initial condition of a narrow diffusing packet (e.g. a Gaussian distribution centered on the origin) and sufficiently long times $t_0$. For power-law initial conditions, a different behavior is expected; see Refs.~\cite{Muk11,Hir11}. For most potentials $U(x)$ (e.g. harmonic), $W(x,t_0)$ in Eq.~\eqref{1} can be replaced by the stationary probability density $W_{\text{eq}}(x)$. For the logarithmic potential, however, it turns out that the integrals in Eq.~\eqref{1} do not always converge and we need to use the explicitly time-dependent probability density $W(x,t_0)$.  In order to solve the Fokker-Planck equation \eqref{A_3a}, we transform it  into  a Schr\"{o}dinger equation and employ  an eigenfunction expansion. Writing the probability density as
$W(x,t) = \chi(x) \psi(x,t)$ with the function $\chi(x) = \exp{(-U(x)/(2 k_B T))}$,
 we obtain the following  equation for $\psi(x,t)$ \cite{Ris86},
\begin{align}
- \frac{1}{D} \frac{\partial}{\partial t} \psi(x,t) = &- \frac{\partial^2}{\partial x^2} \psi(x,t) \nonumber \\
+ \left(\frac{1}{4 k_B^2 T^2} \right. & U'^2(x) + \left.\frac{1}{2 k_B T} U''(x) \right) \psi(x,t) \label{A_6} \ .
\end{align}
Equation \eqref{A_6} has the form of an imaginary-time Schr\"{o}dinger equation with the effective potential,
\begin{align}
U_{\text{eff}}(x)=  \frac{1}{4 k_B^2 T^2} U'^2(x) + \frac{1}{2 k_B T} U''(x)\label{A_7} \ .
\end{align}
Its general solution is given by the expansion,
\begin{align}
\psi(x,t) = \sum_{\lambda} \eta_{\lambda} \psi_{\lambda}(x) \, e^{-D \lambda t} \label{A_8} \ ,
\end{align}
where $\psi_{\lambda}(x) $ are the eigenfunctions of Eq.~\eqref{A_6},
\begin{align}
\left[ -\frac{\partial^2}{\partial x^2} + U_{\text{eff}}(x) \right] \psi_{\lambda}(x) = \lambda \psi_{\lambda}(x) \label{A_8a} \ .
\end{align}
Since the Hamiltonian is symmetric in $x$, we  look for solutions of the eigenvalue equation \eqref{A_8a} that have either even or odd parity \cite{Sha94}. To simplify the calculations, we use the potential
\begin{align}
U(x) = U_0 \ln(|x|) \Theta(|x|-1) \ , \label{A_9}
\end{align}
which has the desired asymptotic form $U(x) \simeq  U_0 \ln |x|$,  $|x|>1$, and is zero for $|x|<1$. The asymptotic long-time behavior of the system is up to a constant factor independent of the potential near the origin and our calculation can be generalized to arbitrary, asymptotically logarithmic potentials \cite{Dec11_2}. For $|x| > 1$, the eigenfunctions are given by  \cite{Mar96}, 
\begin{align}
 \label{A_12}
\psi_{k,e}(x) &= A_k \sqrt{\vert x \vert} \left( a_{1 k} J_{\alpha}(\vert k x \vert) + a_{2 k} J_{-\alpha}(\vert k x \vert) \right) ,\nonumber \\
\psi_{0}(x) &= A \left( a_{1} \vert x \vert^{\frac{1}{2}-\alpha} + a_{2} \vert x \vert^{\alpha + \frac{1}{2}} \right) , \\
\psi_{k,o}(x) &= B_k \, \text{sgn}(k x) \sqrt{\vert x \vert} \left( b_{1 k} J_{\alpha}(\vert k x \vert) + b_{2 k} J_{-\alpha}(\vert k x \vert) \right)  , \nonumber 
\end{align}
with $k = \pm \sqrt{\lambda}$ and
\begin{align}
\alpha = \frac{U_0}{2 k_B T} + \frac{1}{2}  \ .
\end{align}
The subscript $e$ ($o$) refers to the even (odd) solutions,  $J_{\alpha}(x)$ denotes the Bessel function of the first kind, $A$, $A_k$ and $B_k$ are normalization constants and $a_i$, $a_{i k}$ and $b_{i k}$ ($i,j \in \lbrace 1,2 \rbrace$) will be determined by the boundary conditions at $x=\pm 1$. The spectrum for $\alpha > 1$ consists of a single discrete ground state for $k=0$, $\psi_{0}(x)$, which is an even function of $x$, and a continuum of excited states for $k>0$. For $\alpha < 1$, the discrete solution $\psi_{0}(x)$ is non-normalizable (see below) and the spectrum is thus pure continuous. The parameter $\alpha > 1/2$, which measures the ratio of the potential depth $U_0$ and the temperature $T$, will turn out to be the key quantity controlling the long-time behavior of the particle. The structure of the spectrum with the continuum of excited states starting at $k=0$ and thus no gap to the bound state \cite{Far00} is responsible for the anomalous behavior and sensitive dependence on $\alpha$.
Using the eigenstates \eqref{A_12}, the  solution of the Fokker-Planck equation \eqref{A_3a} is
\begin{align}
W(x,t) &= \mathfrak{a}_0 \chi(x) \psi_{0}(x) \nonumber \\
&  + \chi(x) \int_{-\infty}^{\infty} dk \, \mathfrak{a}_k \psi_{k,e}(x) \, e^{-D k^2 t} \nonumber \\
&  + \chi(x) \int_{-\infty}^{\infty} dk \, \mathfrak{b}_k \psi_{k,o}(x) \, e^{-D k^2 t} \label{A_14} \ .
\end{align}
For convenience, we sum over $\pm k$ with $a_{i k} = a_{i (-k)}$ and $b_{i k} = b_{i (-k)}$. The  coefficients $\mathfrak{a}_k$ and $\mathfrak{b}_k$ are determined by the initial condition. 
For $|x|<1$, the potential vanishes and thus the corresponding eigenfunctions are 
\begin{align}
\tilde{\psi}_{k,e}(x) &= A_k \cos(k x) \ ,\nonumber \\
\tilde{\psi}_{0}(x) &= A \ ,\nonumber \\
\tilde{\psi}_{k,o}(x) &= B_k \sin(k x) \label{A_16} \ ,
\end{align}
with the boundary conditions at $x =1$,
\begin{align}
\psi_{k,e}(1) &= \tilde{\psi}_{k,e}(1) \ ,\nonumber \\
\psi_{k,e}'(1) &= \tilde{\psi}_{k,e}'(1) - \frac{U_0}{2 k_B T} \psi_{k,e}(1) \label{A_17} \ .
\end{align}
The last term on the right-hand side  stems from the discontinuity of $U'(x)$ at $x=1$ (similar boundary conditions are obtained for the ground state and the odd eigenfunctions). From this, the coefficients for the ground state are easily seen to be $a_1=1$ and $a_2=0$.

The two integrals in   the solution \eqref{A_14} can be regarded as Laplace transforms. The large $t$ behavior of the system is then determined by the small $k$ expansion of the eigenfunctions, according to the final value theorem \cite{Doe74}. Using the Taylor expansion of the Bessel function  \cite{Abr72}, (Eq.~(9.1.7)), 
\begin{align}
J_{\alpha}(k) &\simeq \frac{1}{\Gamma(\alpha+1)} \left( \frac{k}{2} \right)^{\alpha} \label{A_19} \ ,
\end{align}
and those of the sine and cosine, we find from Eqs.~\eqref{A_12}, \eqref{A_16} and \eqref{A_17} the coefficients of the excited states to leading order in $k$,
\begin{align}
a_{1 k} &\simeq -\Gamma(\alpha) \frac{2\alpha - 1}{\alpha -1} \left( \frac{k}{2} \right)^{2-\alpha} \ ,\nonumber \\
a_{2 k} &\simeq \Gamma(1-\alpha) \left( \frac{k}{2} \right)^{\alpha} \ ,\nonumber \\
b_{1 k} &\simeq \Gamma(\alpha) \left( \frac{k}{2} \right)^{1-\alpha} \ ,\nonumber \\
b_{2 k} &\simeq \Gamma(1-\alpha) \frac{2 \alpha - 1}{\alpha} \left( \frac{k}{2} \right)^{\alpha + 1} \label{A_20} \ .
\end{align}

The probability density $W(x,t)$ should be properly normalized at all times. 
For the discrete ground state, the normalization integral reads,
\begin{align}
2 \int_{0}^{1} dx \, \tilde{\psi}_{0}^2(x) + 2 \int_{1}^{\infty} dx \, {\psi}_{0}^2(x) = 1 \label{A_21} \ .
\end{align}
As it turns out, the ground state is only normalizable for $\alpha > 1$. The corresponding normalization constant is,
\begin{align}
A^2 = \frac{\alpha - 1}{2 \alpha - 1} = \frac{1}{Z} \label{A_22} \ .
\end{align}
Thus the discrete solution is $\psi_0(x) = (1/\sqrt{Z}) \, |x|^{-\alpha + 1/2}$, which is precisely the square root of the normalized stationary solution of the Fokker-Planck equation \eqref{A_3a} for $|x|>1$. For $\alpha < 1$, there is no normalizable ground state and thus no stationary probability density.
Since the excited states form a continuum, they need to be normalized to a delta function,
\begin{align}
2 &\int_{0}^{1} dx \, \tilde{\psi}_{k,e/o}(x) \tilde{\psi}_{k',e/o}(x) \nonumber \\ &+ 2 \int_{1}^{\infty} dx \, {\psi}_{k,e/o}(x) {\psi}_{k',e/o}(x) = \delta(k-k') \ .\label{A_23}
\end{align}
Realizing that the main contribution to the  normalization integral \eqref{A_23} comes from large values of $x$, we use the asymptotic  expansion  of the Bessel function \cite{Abr72}, (Eq.~(9.2.1)),
\begin{align}
J_{\alpha}(k x) \simeq \sqrt{\frac{2}{\pi k x}} \cos \left( k x - \frac{\pi \alpha}{2} - \frac{\pi}{4} \right)  \label{A_24} \ ,
\end{align}
to obtain the expressions, 
\begin{align}
A_k^2 \simeq \frac{k}{4} \frac{1}{a_{1 k}^2 + a_{2 k}^2 + 2 a_{1 k} a_{2 k} \cos \left( 2\phi_{\alpha} -\frac{\pi}{2} \right)}\ , \nonumber \\
B_k^2 \simeq \frac{k}{4} \frac{1}{b_{1 k}^2 + b_{2 k}^2 + 2 b_{1 k} b_{2 k} \cos \left( 2\phi_{\alpha} -\frac{\pi}{2} \right)} \label{A_25}\ ,
\end{align}
where $\phi_{\alpha} = \alpha \pi / 2 + \pi / 4$. Since the coefficients $a_{i k}$ and $b_{j k}$ are only known up to leading order in $k$, we further need to expand the normalization constants $A_k$ and $B_k$ to leading order in $k$ in order to be consistent. Since $\alpha > 1/2$, $b_{1 k}$ will always dominate $b_{2 k}$ and we have
\begin{align}
B_k^2 \simeq \frac{k}{4} \frac{1}{b_{1 k}^2} \label{A_25a}\ .
\end{align}
However, the relative magnitude of $a_{1 k}$ and $a_{2 k}$ depends on the value of $\alpha$. For $\alpha > 1$, $a_{1 k}$ dominates $a_{2 k}$, and vice versa for $\alpha < 1$. This leads to
\begin{align}
A_k^2 \simeq \left\lbrace 
\begin{array}{ll}
\frac{k}{4} \frac{1}{a_{1 k}^2} \quad &\text{for} \quad \alpha > 1 \\[2ex]
\frac{k}{4} \frac{1}{a_{2 k}^2} \quad &\text{for} \quad \alpha < 1 \ .
\end{array} \right.  \label{A_25b} 
\end{align}
The final pieces of information needed to fully determine the solution \eqref{A_14} are the expansion coefficients $\mathfrak{a}_k$ and $\mathfrak{b}_k$. They are related to the initial condition $W(x,0)$ by,
\begin{align}
\mathfrak{a}_{k} &= \int_{-\infty}^{\infty} dx \, \psi_{k,e}(x) \frac{W(x,0)}{\chi(x)}\ , \nonumber \\
\mathfrak{a}_0 &= \int_{-\infty}^{\infty} dx \, \psi_{0}(x) \frac{W(x,0)}{\chi(x)} \ ,\nonumber \\ 
\mathfrak{b}_{k} &= \int_{-\infty}^{\infty} dx \, \psi_{k,o}(x) \frac{W(x,0)}{\chi(x)}\ .  \label{A_26}
\end{align}

\section{Conditional probability density} \label{SEC_3}
The evaluation of the correlation function \eqref{1} requires the computation of the conditional probability density $P(x,t \vert x_0,t_0)$, the solution of the Fokker-Planck equation \eqref{A_3a}  with the initial condition $P(x,t_0)=\delta(x-x_0)$ \cite{Ris86}. Accordingly, it is of the form \eqref{A_14} with the time variable $t$ replaced by the time difference
\begin{align}
\tau = t-t_0 \ .
\end{align}
The expansion coefficients are then, from Eq.~\eqref{A_26},
\begin{align}
\mathfrak{a}_{k} &= \frac{\psi_{k,e}(x_0)}{\chi(x_0)} \ ,\nonumber \\
\mathfrak{a}_0 &= \frac{\psi_{0}(x_0)}{\chi(x_0)} = A \ ,\nonumber \\ 
\mathfrak{b}_{k} &= \frac{\psi_{k,o}(x_0)}{\chi(x_0)}  \label{B_1} \ .
\end{align}
As a result, we have for $x>1$ and $x_0>1$ (the behavior for negative $x$ and $x_0$ follows from symmetry),
\begin{align}
&P(x,t \vert x_0, t_0) = A \chi(x) \psi_{0}(x) \nonumber \\
&\quad + 2 \frac{\chi(x)}{\chi(x_0)} \int_{0}^{\infty} dk \, \psi_{k,e}(x_0) \psi_{k,e}(x) e^{-D k^2 \tau} \nonumber \\
&\quad + 2 \frac{\chi(x)}{\chi(x_0)} \int_{0}^{\infty} dk \, \psi_{k,o}(x_0) \psi_{k,o}(x) e^{-D k^2 \tau} \nonumber \\
&\simeq A^2 x^{1-2\alpha} + 2 \, x^{1-\alpha} x_0^{\alpha} \int_{0}^{\infty} dk \nonumber \\
&\quad \times \Big[ A_k^2 \left(a_{1 k} J_{\alpha}(k x_0) + a_{2 k} J_{-\alpha}(k x_0)\right) \nonumber \\
&\qquad \times \left(a_{1 k} J_{\alpha}(k x) + a_{2 k} J_{-\alpha}(k x)\right) e^{-D k^2 \tau} \nonumber \\
&\qquad + B_k^2 \left(b_{1 k} J_{\alpha}(k x_0) + b_{2 k} J_{-\alpha}(k x_0)\right) \nonumber \\
&\qquad \times \left(b_{1 k} J_{\alpha}(k x) + b_{2 k} J_{-\alpha}(k x)\right) e^{-D k^2 \tau} \Big] \label{B_2} \ .
\end{align}
For completeness, the discussion of the cases where either $x$ or $x_0$  is smaller than $1$ is given in Appendix \ref{SEC_A}. The very first term on the right-hand side of Eq.~\eqref{B_2} is the contribution of the stationary state and only appears for $\alpha > 1$ (since $A = 0$ for $\alpha < 1$). The expansion of the product in the integral yields four terms for the even contributions (those containing the $a_{i k}$-coefficients) and four terms for the odd contributions (those containing the $b_{j k}$-coefficients). Since the leading order of the normalization constant $A_k$ is different depending on whether $\alpha$ is larger or smaller than $1$ (see Eq.~\eqref{A_25b}), we have to examine 12 integrals in total (some of which are fortunately the same). However, all these integrals share the same basic structure:
\begin{align}
\mathfrak{I}_{e/o,\mu,\nu}(z,y) &= c_{e/o,\mu,\nu}(z,y) \nonumber \\
& \, \times \int_{0}^{\infty} du \, u^{\lambda_{e/o,\mu,\nu}} J_{\mu} (2 u y) J_{\nu} (2 u z) e^{-u^2} \label{B_3} \ ,
\end{align}
where $\mu$ and $\nu$ are equal to $\pm \alpha$ and both the $u$-independent factor $c_{e/o,\mu,\nu}(z,y)$ and the exponent $\lambda_{e/o,\mu,\nu}$ are different depending on the combination of $\mu$ and $\nu$. Here we have introduced the variables $u=k(D \tau)^{1/2}$, $z=x/(4 D \tau)^{1/2}$ and $y=x_0/(4 D \tau)^{1/2}$. We can then write Eq.~\eqref{B_2} as,
\begin{align}
P(x&,t \vert x_0,t_0) \simeq A^2 x^{-2\alpha + 1} \nonumber \\  
& \, + \sum_{e,o} \sum_{\mu,\nu = \pm \alpha} \, \left(\sqrt{D \tau}\right)^{2-\lambda_{e/o,\mu,\nu}} \mathfrak{I}_{e/o,\mu,\nu}(z,y) \label{B_4} \ .
\end{align}
It is important to note that the main contribution to the integrals $\mathfrak{I}_{e/o,\mu,\nu}(z,y)$ comes from small values of $u$ as the exponential factor causes the integrand to vanish exponentially for large values of $u$ and all other functions increase at most as a power of $u$. Table \ref{TAB_1} gives a summary of all  the values of $\mu$, $\nu$, and $\lambda_{e/o,\mu,\nu}$, as well as the prefactors $c_{e/o,\mu,\nu}(z,y)$ of the respective integrals.

\begin{table*}
\begin{align}
\begin{array}{|l|c|c|c|c|}
\hline & \text{prefactor} \, c_{e/o,\mu,\nu}(z,y) & \lambda_{e/o,\mu,\nu} & \mu & \nu  \\ 
\hline \text{even contribution}, \; \alpha > 1 & y^{\alpha} z^{-\alpha+1} \times \left \lbrace \begin{array}{lr} 1 \\[1 ex] \frac{2^{2-2\alpha}\Gamma(2-\alpha)}{\Gamma(\alpha)(2\alpha-1)} \\[1 ex] \frac{2^{2-2\alpha}\Gamma(2-\alpha)}{\Gamma(\alpha)(2\alpha-1)} \\[1 ex] \left( \frac{2^{2-2\alpha}\Gamma(2-\alpha)}{\Gamma(\alpha)(2\alpha-1)} \right)^2 \end{array} \right. & \begin{array}{lr} 1 \\[1 ex] 2\alpha-1 \\[1 ex] 2\alpha-1 \\[1 ex] 4\alpha-3 \end{array} & \begin{array}{lr} \alpha \\[1 ex] \alpha \\[1 ex] -\alpha \\[1 ex] -\alpha \end{array} & \begin{array}{lr} \alpha \\[1 ex] -\alpha \\[1 ex] \alpha \\[1 ex] -\alpha \end{array} \\ [8 ex]
\hline \text{even contribution}, \; \alpha < 1 & y^{\alpha} z^{-\alpha+1} \times \left \lbrace \begin{array}{lr}  \left( \frac{2^{2\alpha-2}\Gamma(\alpha)(2\alpha-1)}{\Gamma(2-\alpha)} \right)^2 \\[1 ex] \frac{2^{2\alpha-2}\Gamma(\alpha)(2\alpha-1)}{\Gamma(2-\alpha)} \\[1 ex] \frac{2^{2\alpha-2}\Gamma(\alpha)(2\alpha-1)}{\Gamma(2-\alpha)} \\[1 ex] 1 \end{array} \right. & \begin{array}{lr} -4\alpha+5 \\[1 ex] -2\alpha+3 \\[1 ex] -2\alpha+3 \\[1 ex] 1  \end{array} & \begin{array}{lr} \alpha \\[1 ex] \alpha \\[1 ex] -\alpha \\[1 ex] -\alpha \end{array} & \begin{array}{lr} \alpha \\[1 ex] -\alpha \\[1 ex] \alpha \\[1 ex] -\alpha \end{array} \\ [8 ex]
\hline \text{odd contribution}& y^{\alpha} z^{-\alpha+1} \times \left \lbrace \begin{array}{lr} 1 \\[1 ex] \frac{2^{-2\alpha}\Gamma(1-\alpha)(2\alpha-1)}{\Gamma(\alpha+1)} \\[1 ex] \frac{2^{-2\alpha}\Gamma(1-\alpha)(2\alpha-1)}{\Gamma(\alpha+1)} \\[1 ex] \left( \frac{2^{-2\alpha}\Gamma(1-\alpha)(2\alpha-1)}{\Gamma(\alpha+1)} \right)^2 \end{array} \right. & \begin{array}{lr} 1 \\[1 ex] 2\alpha+1 \\[1 ex] 2\alpha+1 \\[1 ex] 4\alpha+1 \end{array} & \begin{array}{lr} \alpha \\[1 ex] \alpha \\[1 ex] -\alpha \\[1 ex] -\alpha \end{array} & \begin{array}{lr} \alpha \\[1 ex] -\alpha \\[1 ex] \alpha \\[1 ex] -\alpha \end{array} \\ [8 ex]
\hline
\end{array} \nonumber
\end{align}
\caption{Prefactors and parameters for the integrals appearing in the conditional probability density, Eq.~\eqref{B_4}.}
\label{TAB_1}
\end{table*}

\subsection{Probability density function} \label{SEC_3A}
For very large times and fixed $x_0$, the conditional probability density loses its dependence on the initial condition and reduces to the probability density, $P(x,t \vert x_0,t_0)\simeq  W(x,\tau)$ [see Eq.~\eqref{B_6} below]. In this limit, 
the variable $y=x_0/(4 D \tau)^{1/2}$  is small, and the Bessel function containing $y$ in Eq.~\eqref{B_4} can be expanded using Eq.~\eqref{A_19} to get
\begin{align}
\label{B_5}
\mathfrak{I}_{e/o,\mu,\nu}(z,y) &\simeq \frac{1}{\Gamma(\mu + 1)} c_{e/o,\mu,\nu}(z,y) \, y^{\mu} \nonumber \\
& \, \times \int_{0}^{\infty} du \, u^{\lambda_{e/o,\mu,\nu} + \mu} J_{\nu} (2 u z) e^{-u^2} \ .
\end{align}
Note that the variable $z=x/(4 D \tau)^{1/2}$ is not necessarily small, since we are  interested in the large $x$ regime. Equation~\eqref{B_5} allows us to find the leading-order contribution to $W(x,\tau)$ at very large times without having to  compute explicitly all the integrals. Comparing the values for $\mu$ and $\lambda_{e/o,\mu,\nu}$ for the different integrals, one finds that the leading contribution stems from the even terms with $\mu = -\alpha$ and $\nu = \alpha$ for $\alpha >1$, and from the even terms containing $\mu = -\alpha$ and $\nu = -\alpha$ for $\alpha <1$. Evaluating the respective integrals using Eq.~(6.631.1) of Ref.~\cite{Gra07}, we obtain the asymptotic behavior of the probability density for $x > 1$,
\begin{align}
W(x,\tau) \simeq \left\lbrace
\begin{array}{ll}
\frac{1}{Z \Gamma(\alpha)} x^{1-2 \alpha} \Gamma \left( \alpha,\frac{x^2}{4 D \tau} \right) \; &\text{for} \; \alpha > 1 \\[2ex]
\frac{1}{\Gamma(1-\alpha)} (4 D \tau)^{\alpha - 1} x^{1-2\alpha} e^{-\frac{x^2}{4 D \tau}} \; &\text{for} \; \alpha < 1 .
\end{array} \right. \label{B_6} 
\end{align}
Here $\Gamma(\alpha,x)$ is the incomplete Gamma function. Equation \eqref{B_6} for $\alpha > 1$ is the infinite covariant density (ICD) \cite{Bar10,Dec11_2,Hir11} .
This result depends on the specific form of the potential $U(x)$ only through the partition function $Z$, as long as it is regular at the origin and has the same asymptotically logarithmic behavior as Eq.~\eqref{A_9}. 
The ICD is non-normalizable, but allows us to compute the asymptotic long-time behavior of the moments $\langle | x |^q \rangle$ of order $q > 2 \alpha - 2$ \cite{Bar10}. The lower-order moments are finite for the equilibrium density $W_{\text{eq}}(x) = e^{-U(x)/(k_B T)}/Z$ and thus can be obtained from the latter. For $\alpha < 1$, the asymptotic form \eqref{B_6} of the probability density is normalizable and can be used to calculate all moments. 
The asymptotic behavior of  the second moment  is given by
\begin{align}
&\langle x^2(\tau) \rangle = 2 \int_{0}^{\infty} dx \, x^2W(x,\tau) \nonumber \\
&\simeq \left\lbrace
\begin{array}{ll}
\frac{\alpha - 1}{3 \alpha - 6}  \; &\text{for} \; \alpha > 2 \\[2ex]
\frac{1}{Z\Gamma(\alpha) (2-\alpha)} (4 D \tau)^{2-\alpha} \; &\text{for} \; 1 < \alpha < 2 \\[2ex] (1-\alpha) 4 D \tau \; &\text{for} \; \alpha < 1 \ .
\end{array} \right. \label{B_7}
\end{align}
For $\alpha > 2$, the second moment tends to a constant which depends explicitly on the specific shape of the potential for $x$ of $\mathcal{O}(1)$ and can be obtained from the equilibrium distribution.  For $\alpha < 2$, the second moment increases with time and diverges in the infinite time limit. In the range $1 < \alpha < 2$, we observe subdiffusive behavior, which depends on the shape of the potential for $x$ of $\mathcal{O}(1)$ only through the partition function $Z$. When $\alpha < 1$, the diffusion is normal and asymptotically independent of the potential at small $x$. 
As it turns out, the ICD Eq.~\eqref{B_6} not only governs the time dependence of the higher-order moments, but is also crucial for the calculation of the correlation function (see section \ref{SEC_4}).

\subsection{Conditional probability density} \label{SEC_3B}
In order to calculate the conditional probability density, we consider times long enough that we can use the small-$k$ expansion of the coefficients, while not necessarily so long that we can ignore the initial condition. In this regime neither $z=x/(4 D \tau)^{1/2}$ nor $y=x_0/(4 D \tau)^{1/2}$ appearing in Eq.~\eqref{B_4} is small. In the case of both $z$ and $y$ (and thus $x$ and $x_0$) being of the same order, not all the integrals can be explicitly evaluated. However, we can estimate the term which is of leading order for large times by evaluating the integrals for $z = y$.
Fortunately, the integrals for the leading order term can be computed explicitly. For both $z$ and $y$ at least of order unity, the leading order term for long times is then [see Ref.~\cite{Gra07}, Eq.~(6.633.2) for the computation of the integrals], 
\begin{align}
P_{e}&(x,t \vert x_0, t_0) \nonumber \\
&\simeq \left\lbrace
\begin{array}{ll}
\frac{1}{Z} x^{1-2 \alpha} + (4 D \tau)^{-1} x^{1-\alpha} x_0^{\alpha} \\
 \quad \times \exp \left(-\frac{x^2 + x_0^2}{4 D \tau}\right) I_{\alpha} \left( \frac{x x_0}{2 D \tau} \right)  \quad &\text{for} \; \alpha > 1 \\[2ex]
(4 D \tau)^{-1} x^{1-\alpha} x_0^{\alpha} \\
 \quad \times \exp \left(-\frac{x^2 + x_0^2}{4 D \tau}\right) I_{-\alpha} \left( \frac{x x_0}{2 D \tau} \right) \quad &\text{for} \; \alpha < 1
\end{array} \right. \label{B_8}
\end{align}
for the even contribution, $I_\alpha(x)$ is the modified Bessel function of the first kind, and 
\begin{align}
P_{o}&(x,t \vert x_0, t_0) \nonumber \\ 
&\simeq (4 D \tau)^{-1} x^{1-\alpha} x_0^{\alpha} \exp \left(-\frac{x^2 + x_0^2}{4 D \tau}\right) I_{\alpha} \left( \frac{x x_0}{2 D \tau} \right) \label{B_9}
\end{align}
for the odd contribution. The latter is the same as the probability density for the Bessel process with an absorbing boundary condition at the origin \cite{Bra00,Mar11}. The Bessel process describes diffusion in a purely logarithmic potential (and thus a $1/x$-force) which is not regular at the origin. While the Bessel process correctly describes the asymptotic behavior of the odd moments, its Green's function is well defined only for $\alpha < 1$, when the singularity on the origin is integrable. Thus for the calculation of the even moments (including normalization) for $\alpha > 1$, the regularization of the potential on the origin is vital, since only then do we have a finite partition function $Z$. The odd part of the probability density does not depend on $Z$ at all and is identical to that of the Bessel process.

If one of the variables $z$ and $y$ is much bigger than the other, the Bessel function containing this variable will oscillate rapidly (see e.g. Eq.~\eqref{A_24}) for all but very small values of the variable of integration $u$. We may then expand the Bessel function containing the other variable for small arguments. For the leading order even contribution, we obtain the ICD Eq.~\eqref{B_6} if $z \gg y$ ($x \gg x_0$), and
\begin{align}
P_{e}&(x,t \vert x_0, t_0) \nonumber \\
& \simeq \left\lbrace
\begin{array}{ll}
\frac{1}{Z \Gamma(\alpha)} x^{1-2 \alpha} \Gamma \left( \alpha,\frac{x_0^2}{4 D \tau} \right) \quad &\text{for} \; \alpha > 1 \\[2ex]
\frac{1}{\Gamma(1-\alpha)} (4 D \tau)^{\alpha - 1} x^{1-2\alpha} e^{-\frac{x_0^2}{4 D \tau}} \quad &\text{for} \; \alpha < 1
\end{array} \right. \label{B_10}
\end{align}
if $y \gg z$ ($x_0 \gg x$). Note that Eqs.~\eqref{B_6} and \eqref{B_10} are not limiting forms of Eq.~\eqref{B_8}, since the leading order term is different depending on the size of $z$ and $y$. The leading order odd contribution is always given by Eq.~\eqref{B_9}.
Using the odd part of the conditional probability density \eqref{B_9}, we can directly calculate the average $E(x_0,\tau)$ of the position $x$ (for $x_0 > 1$),
\begin{align}
&E(x_0,\tau) = 2 \int_{0}^{\infty} dx \, x P_o(x,t \vert x_0,t_0) \nonumber \\
&\simeq \frac{\sqrt{\pi}}{2 \Gamma(\alpha+1)} (4 D \tau)^{\frac{1}{2}-\alpha} x_0^{2 \alpha} e^{-\frac{x_0^2}{4 D t}} {}_{1}\text{F}_{1} \left( \frac{3}{2}; \alpha + 1; \frac{x_0^2}{4 D \tau} \right) . \label{B_11}
\end{align}
Here ${}_{1}\text{F}_{1} (a; b; x)$ is the confluent hypergeometric function. For long times, $4 D \tau \gg x_0^2$, the hypergeometric function is approximately unity, and we thus find,
\begin{align}
E(x_0, &\tau) \simeq \frac{\sqrt{\pi}}{2 \Gamma(\alpha+1)} (4 D \tau)^{\frac{1}{2}-\alpha} x_0^{2 \alpha} \ . \label{B_12}
\end{align}
For $\alpha = 1/2$ ($U_0 = 0$), we recover $E(x_0,\tau) = x_0 $, which corresponds to free diffusion. For $\alpha > 1/2$, on the other hand, we observe an algebraic dependence on the initial position $x_0$.

\section{Correlation function} \label{SEC_4}
\begin{figure}
\includegraphics[trim=20mm 5mm 25mm 15mm, clip, width=0.47\textwidth]{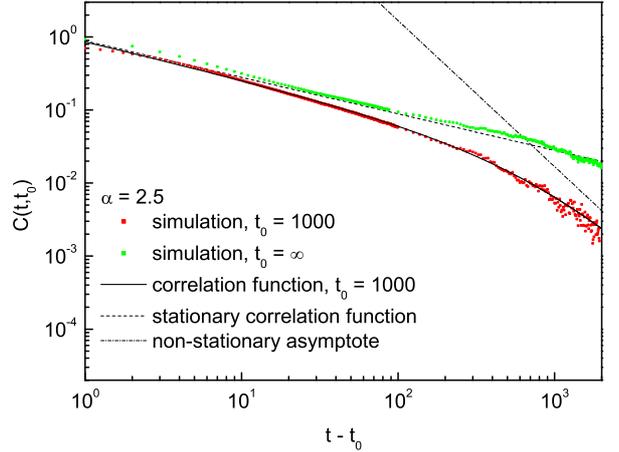}
\caption{(Color online) The correlation function $C(t,t_0)$  for $U(x) = (U_0/2) \ln(1+x^2)$ (thus $Z=\sqrt{\pi} \Gamma(\alpha-1)/\Gamma(\alpha-1/2)$) and $\alpha = 2.5$ ($U_0 = 1$, $k_B T = 0.25$, $\gamma = 1$). The solid black line is the analytical expression \eqref{C_6}. The  red (dark gray) and  green (light grey) dots are the results of Langevin simulations for two different times, $t_0 = 1000$ and $t_0 =\infty$ (i.e. starting from the equilibrium distribution). For $t_0 =1000$, we observe a transition from the stationary behavior Eq.~\eqref{C_9} to the aging form Eq.~\eqref{C_8}.  For $t_0 = \infty$, we observe the stationary correlation  function \eqref{C_9} at all times $t$.}
\label{Fig_4_1}
\end{figure}

Having derived the asymptotic behavior of  the conditional probability density and of the probability density function, we can next  evaluate the correlation function \eqref{1}. Since the probability density $W(x_0,t_0)$, as given by  Eq.~\eqref{B_6}, is an even function of $x_0$, only the odd part of the conditional probability density contributes to the integral of the correlation function:
\begin{align}
&C(t,t_0) \simeq 4 \int_{1}^{\infty} dx \int_{1}^{\infty} dx_0 \, x x_0 \underbrace{P_{o}(x,t \vert x_0,t_0)}_{\text{Bessel process}} \underbrace{W(x_0,t_0)}_{\text{ICD}} \label{C_0}
\end{align}
This result is remarkably intuitive: For $\alpha > 1$, the infinite covariant density (ICD) gives the probability for finding the particle at $x_0$, while the Bessel process describes the relaxation of the particle's average position from $x_0$ toward the origin. Since we use the asymptotic expressions for the conditional probability density and the probability density, the resulting expression for the correlation function is also asymptotic in the sense that we require both $\tau = t-t_0$ and $t_0$ to be large. Inserting the results Eqs.~\eqref{B_6} and \eqref{B_9} into Eq.~\eqref{C_0}, we obtain for $\alpha >1$,
\begin{align}
C&(t,t_0) \nonumber \\
&\; \simeq \frac{4}{Z \Gamma(\alpha)} (4 D \tau)^{-1} \int_{1}^{\infty} dx \int_{1}^{\infty} dx_0 \, x^{2-\alpha} x_0^{2-\alpha} \nonumber \\
& \quad \times \exp \left(-\frac{x^2 + x_0^2}{4 D \tau}\right) I_{\alpha} \left( \frac{x x_0}{2 D \tau} \right) \Gamma \left( \alpha,\frac{x_0^2}{4 D t_0} \right) \label{C_1} \ .
\end{align}
With the help of the variables $z=x/(4 D \tau)^{1/2}$ and $y=x_0/(4 D \tau)^{1/2}$ introduced above, Eq.~\eqref{C_1} simplifies to
\begin{align}
C&(t,t_0) \simeq \frac{4}{Z \Gamma(\alpha)} (4 D \tau)^{2-\alpha} \nonumber \\
&\times \int_{\frac{1}{\sqrt{4 D \tau}}}^{\infty} dz \int_{\frac{1}{\sqrt{4 D \tau}}}^{\infty} dy \, z^{2-\alpha} y^{2-\alpha} \nonumber \\
& \quad \times e^{-(z^2 + y^2)} I_{\alpha} \left( 2 z y \right) \Gamma \left( \alpha, y^2 \frac{\tau}{t_0} \right) \label{C_2} \ .
\end{align}
For long times, $\tau \gg 1$, we may take the lower boundary of the integrals to $0$, since the integrand vanishes as $z^2$ and $y^2$ [note that $I_{\alpha}(x) \simeq (x/2)^{\alpha}/\Gamma(\alpha + 1) $ when $x \rightarrow 0$]. The integral over $z$ can then be evaluated using Eq.~(6.631.1) of Ref.~\cite{Gra07}, together with $I_{\alpha}(x)=(-i)^{\alpha} J_{\alpha} (i x)$, to give,
\begin{align}
C&(t,t_0) \simeq \frac{\sqrt{\pi} }{Z \Gamma(\alpha + 1) \Gamma(\alpha)} (4 D \tau)^{2-\alpha} \nonumber \\
&\times \int_{0}^{\infty} dy \, y^{2} e^{-y^2} \, {}_{1}\text{F}_{1} \left( \frac{3}{2}; \alpha + 1; y^2 \right) \Gamma \left( \alpha, y^2 \frac{\tau}{t_0} \right) \label{C_3} \ .
\end{align}
At the same time, for $\alpha < 1$, we have the integral,
\begin{align}
C&(t,t_0) \simeq \frac{4}{\Gamma(1 - \alpha)} (4 D \tau)^{2-\alpha} (4 D t_0)^{\alpha - 1} \nonumber \\
&\times \int_{\frac{1}{\sqrt{4 D \tau}}}^{\infty} dz \int_{\frac{1}{\sqrt{4 D \tau}}}^{\infty} dy \, z^{2-\alpha} y^{2-\alpha} \nonumber \\
& \quad \times e^{-(z^2 + y^2)} I_{\alpha} \left( 2 z y \right) e^{-y^2 \frac{\tau}{t_0}} \label{C_4} \ .
\end{align}
For large $\tau$, the latter expression reduces to  
\begin{align}
C&(t,t_0) \simeq \frac{\sqrt{\pi}}{\Gamma(\alpha + 1) \Gamma(1 - \alpha)} (4 D \tau) \left( \frac{\tau}{t_0} \right)^{1-\alpha} \nonumber \\
&\times \int_{0}^{\infty} dy \, y^{2} e^{-y^2} \, {}_{1}\text{F}_{1} \left( \frac{3}{2}; \alpha + 1; y^2 \right) e^{-y^2 \frac{\tau}{t_0}} \ . \label{C_5}
\end{align}
Using $\tau = t-t_0$, the structure of the results \eqref{C_3} and \eqref{C_5} can be summarized in the compact form:
\begin{align}
C(t,t_0) \simeq \left\lbrace 
\begin{array}{ll}
\frac{\sqrt{\pi}}{Z \, \Gamma(\alpha + 1) \Gamma(\alpha)} \, (4 D (t-t_0))^{2-\alpha} f_{\alpha} \left( \frac{t-t_0}{t_0} \right) \\ 
\qquad \text{for} \; \alpha > 1 \\[2ex]
\frac{\sqrt{\pi}}{\Gamma(\alpha + 1) \Gamma(1 - \alpha)} \, 4 D (t-t_0) \, g_{\alpha} \left( \frac{t-t_0}{t_0} \right) \\
\qquad \text{for} \; \alpha < 1
\end{array} \right.  \label{C_6}
\end{align}
where we have introduced the two functions,
\begin{align}
f_{\alpha}(s) &=  \int_{0}^{\infty} dy \, y^{2} e^{-y^2} \, {}_{1}\text{F}_{1} \left( \frac{3}{2}; \alpha + 1; y^2 \right) \Gamma \left( \alpha, y^2 s \right)\ , \nonumber \\
g_{\alpha}(s) &=  s^{1-\alpha} \int_{0}^{\infty} dy \, y^{2} e^{-y^2} \, {}_{1}\text{F}_{1} \left( \frac{3}{2}; \alpha + 1; y^2 \right) e^{-y^2 s}\label{C_7}\ .
\end{align}
Note that the correlation function depends on the specific shape of the potential for $|x|<1$ only via the partition function $Z$ (for $\alpha > 1$) or not at all (for $\alpha < 1$). These results are thus valid for arbitrary (regular) potentials with the same asymptotic behavior as Eq.~\eqref{A_9}.  The behaviors of  $f_{\alpha}(s)$ and $g_{\alpha}(s)$ for small and large arguments (corresponding to $t_0 \gg t-t_0$ and $t-t_0 \gg t_0$) are given  in Appendix \ref{SEC_D}. For  $t-t_0 \gg t_0$, we obtain,
\begin{align}
C(t,t_0) \simeq \left\lbrace 
\begin{array}{ll}
\frac{\sqrt{\pi} \Gamma\left(\alpha + \frac{3}{2}\right)}{3 Z \, \Gamma(\alpha + 1) \Gamma(\alpha)} \\
\; \times \left(\frac{t-t_0}{t_0}\right)^{\frac{1}{2}-\alpha} (4 D t_0)^{2-\alpha} \; &\text{for} \; \alpha > 1 \\[2ex]
\frac{\pi}{4 \Gamma(\alpha + 1) \Gamma(1 - \alpha)} \\
\; \times \left(\frac{t-t_0}{t_0}\right)^{\frac{1}{2}-\alpha} (4 D t_0) \; &\text{for} \; \alpha < 1 \ .
\end{array} \right.  \label{C_8}
\end{align}
In both cases, the correlation function is nonstationary, its value increases with the initial time $t_0$ and decays as $(t-t_0)^{1/2-\alpha}$. We note that this is the same  time-dependence as  the first moment $E(x_0,\tau)$, Eq.~\eqref{B_12}. For $t_0 \gg t-t_0$, on the other hand, we find,
\begin{align}
C&(t,t_0) \simeq \nonumber \\
&\left\lbrace 
\begin{array}{ll}
\frac{\pi \, \Gamma(\alpha - 2) }{4 Z \, \Gamma^2 \left( \alpha - \frac{1}{2} \right) } (4 D (t - t_0))^{2-\alpha} \; &\text{for} \; \alpha > 2 \\[2ex]
\frac{1}{Z\, \Gamma(\alpha)(2-\alpha)} (4 D t_0)^{2-\alpha} \; &\text{for} \; 1 < \alpha < 2 \\[2ex]
(1-\alpha) 4 D t_0 \; &\text{for} \; \alpha < 1 \ .
\end{array} \right.  \label{C_9}
\end{align}
For $\alpha>2$, the correlation function is stationary in this limit. This is illustrated in Fig.~\ref{Fig_4_1}: For times $t-t_0$ that are short compared to $t_0$, we observe the stationary behavior Eq.~\eqref{C_9}. For longer times, there is a transition to the nonstationary behavior Eq.~\eqref{C_8}, though this is difficult to observe in our Langevin simulations. If we start out in  the stationary state (which corresponds to $t_0 = \infty$), we have the stationary behavior Eq.~\eqref{C_9} at all times for $\alpha > 2$.

\begin{figure}
\includegraphics[trim=20mm 10mm 25mm 15mm, clip, width=0.47\textwidth]{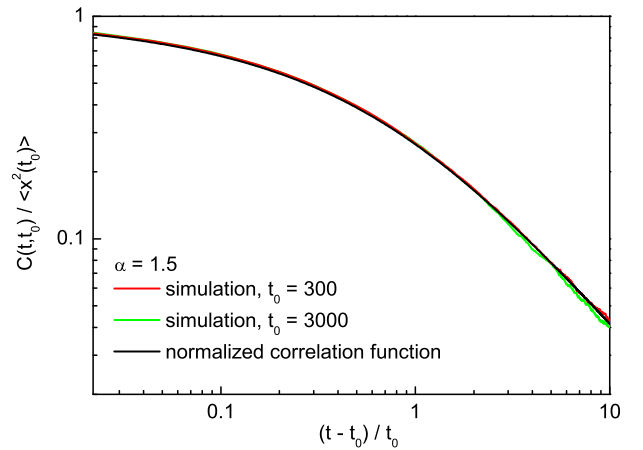}
\caption{(Color online) The normalized correlation function $C(t,t_0)/\langle x^2(t_0)\rangle$  for $U(x) = (U_0/2) \ln(1+x^2)$ and $\alpha = 1.5$ ($U_0=1$, $k_B T = 0.5$, $\gamma = 1$). The solid black line is the analytical result \eqref{C_10}. The numerical simulations  for $t_0= 300$ (red/dark gray) and $t_0=3000$ (green/light gray) perfectly confirm the super-aging behavior.}
\label{Fig_4_2}
\end{figure}

For $\alpha < 2$, the asymptotic behavior of the correlation function is nonstationary and is dominated by the increase of the second moment  $\langle x^2(t_0)\rangle$, Eq.~\eqref{B_7}, which means that using the stationary distribution $W_{\text{eq}}(x)$ to calculate the correlation function gives an infinite value. In this case, the correlation function can be expressed in a convenient way by normalizing it to the value of the second moment $\langle x^2(t_0)\rangle$:
\begin{align}
\frac{C(t,t_0)}{\langle x^2(t_0)\rangle} \simeq & \left\lbrace
\begin{array}{ll}
\frac{\sqrt{\pi}(2 - \alpha)}{\Gamma(\alpha + 1)} \, \tilde{f}_{\alpha} \left( \frac{t-t_0}{t_0} \right) \; &\text{for} \; 1< \alpha < 2 \\ [2ex]
\frac{\sqrt{\pi}}{\Gamma(\alpha + 1) \Gamma(2 - \alpha)} \, \tilde{g}_{\alpha} \left( \frac{t-t_0}{t_0} \right) \; &\text{for} \; \alpha < 1
\end{array} \right. \label{C_10}
\end{align}
with
\begin{align}
\tilde{f}_{\alpha}(s) &=  s^{2-\alpha} f_{\alpha}(s) \ , \nonumber \\
\tilde{g}_{\alpha}(s) &= s  g_{\alpha}(s) \ . \label{C_11}
\end{align}
In this regime, the system exhibits aging. However, contrary to usual aging behavior, which is of the form $C(t,t_0) = \langle x^2\rangle_\text{eq} f((t-t_0)/t_0) $ \cite{Bou92,Bur10,Mar04}, Eq.~\eqref{C_11} shows that here $C(t,t_0) = \langle x^2(t_0)\rangle f((t-t_0)/t_0) $. Since the prefactor increases with time, we call this behavior super-aging. A similar behavior, albeit with a logarithmic time dependence, has been observed for Sinai's model for diffusion in a random environment \cite{Fis99}. The stationary correlation function for $\alpha > 2$ \eqref{C_9} agrees with the result derived  using the equilibrium  solution of the Fokker-Planck equation \cite{Mar96}. However, neither the super-aging behavior nor the long-time limit Eq.~\eqref{C_8} can be obtained from the  equilibrium distribution: both require the infinite covariant density \eqref{B_6}.
 Figure \ref{Fig_4_2} shows the normalized correlation function \eqref{C_10} for different values of $t_0$. It clearly illustrates the super-aging behavior of the correlation function.

\section{Variance of the time average} \label{SEC_5}
We are now in a position to evaluate the long-time behavior of the variance $\bar{\sigma}^2(t)$ of the time-averaged position \eqref{2},
\begin{align}
\bar{\sigma}^2(t) &= \langle \bar{x}^2(t) \rangle - \langle \bar{x}(t) \rangle^2 \ .
\end{align}
We begin with the second term,
\begin{align}
\langle \bar{x}(t) \rangle^2 = \left( \frac{1}{t-t^*} \int_{t^*}^{t} dt' E(x_0,t') \right)^2 \ ,
\end{align}
 involving the integral over $E(x_0,t) = \langle x(x_0,t) \rangle$. Taking the time average of the first moment \eqref{B_12}, we obtain,
\begin{align}
\langle \bar{x}(t) \rangle &\simeq \frac{\sqrt{\pi}}{3 - 2 \alpha} \frac{1}{4 D (t-t^{*})} x_0^{2 \alpha} \left[ (4 D t')^{\frac{3}{2}-\alpha}\right]_{t^{*}}^{t} \ .\label{D_3}
\end{align}
Since we only know the asymptotic long-time behavior of quantities like the first moment $E(x_0,t)$ and the correlation function $C(t,t_0)$ [the short time behavior will in general depend on the shape of the potential $U(x)$ for small $x$], we calculate the time average starting at some time $t^{*}$, which we assume to be large enough that the asymptotic expressions are valid. For large times, $t \gg t^{*}$, we then have,
\begin{align}
&\langle \bar{x}(t) \rangle^2 \nonumber \\
&\simeq \left\lbrace 
\begin{array}{ll}
\frac{\pi}{(3 - 2 \alpha)^2} x_0^{4 \alpha} (4 D t)^{1-2\alpha} \; &\text{for} \; \alpha < \frac{3}{2} \\[2ex]
\frac{\pi}{(2 \alpha - 3)^2} x_0^{4 \alpha} (4 D t)^{-2} (4 D t^{*})^{3-2\alpha} \; &\text{for} \; \alpha > \frac{3}{2} \ .
\end{array} \right. \label{D_4}
\end{align}
The analysis of the first term,
\begin{align}
\langle \bar{x}^2(t) \rangle = \frac{1}{(t-t^*)^2} \int_{t^*}^{t} dt'' \int_{t^*}^{t} dt' C(t'',t') \ , \label{cint}
\end{align}
is carried out in detail in Appendix \ref{SEC_C}. We obtain
\begin{align}
&\bar{\sigma}^2(t) \simeq \langle \bar{x}^2(t) \rangle \simeq c_{\alpha} \left\lbrace 
\begin{array}{ll}
(4 D t)^{-1} & \text{for} \quad \alpha > 3 \\[2ex]
(4 D t)^{2-\alpha} &\text{for} \quad 1 < \alpha < 3 \\[2 ex]
4 D t &\text{for} \quad \alpha < 1 \ ,
\end{array} \right. \label{D_13}
\end{align}
with
\begin{align}
c_{\alpha} = \left\lbrace
\begin{array}{ll}
\frac{\sqrt{\pi}}{Z \, \Gamma(\alpha+1) \Gamma(\alpha)(4 - \alpha)} \, \int_{0}^{\infty} ds \, \frac{s^{2-\alpha}}{(s+1)^{4-\alpha}} \, f_{\alpha}(s) \\
\qquad \qquad \; \text{for} \quad 1 < \alpha < 3 \\[2 ex]
\frac{\sqrt{\pi}}{3 \Gamma(\alpha) \Gamma(1-\alpha)} \, \int_{0}^{\infty} ds \, \frac{s}{(s+1)^{3}} \, g_{\alpha}(s) \\
\qquad \qquad \; \text{for} \quad \alpha < 1 \ ,
\end{array} \right. \label{D_13a}
\end{align}
which dominates the contribution from Eq.~\eqref{D_4} for long times and thus determines the behavior of $\bar{\sigma}^2(t)$. The prefactor $c_{\alpha}$ for $\alpha > 3$ depends on the specific choice of the potential (see below). Note that the asymptotic result for $\bar{\sigma}^2(t)$ is independent of $t^*$, as it should be.

\begin{figure}
\includegraphics[trim=20mm 5mm 25mm 15mm, clip, width=0.47\textwidth]{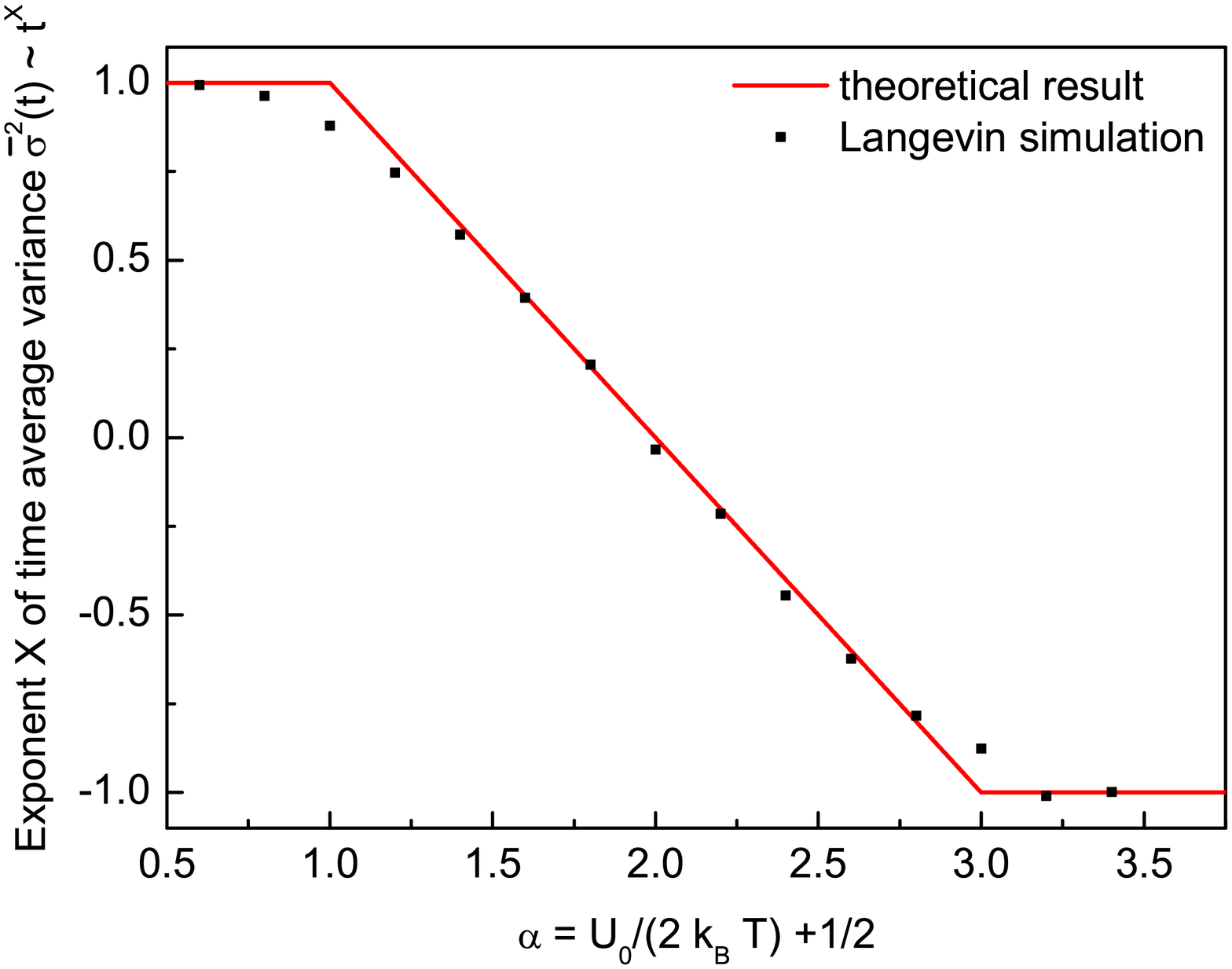}
\caption{(Color online) Power-law exponent of the long-time variance $\bar{\sigma}^2(t) \simeq t^X$, Eq.~\eqref{D_13} (red solid line), and numerical data obtained from Langevin simulations by fitting the long-time behavior up to $t=4000$ (black dots), as a function of the parameter $\alpha= U_0/(2 k_B T) + 1/2$, for the potential $U(x) = (U_0/2) \ln(1+x^2)$.}
\label{Fig_5_1}
\end{figure}

\begin{figure}
\includegraphics[trim=18mm 10mm 25mm 15mm, clip, width=0.47\textwidth]{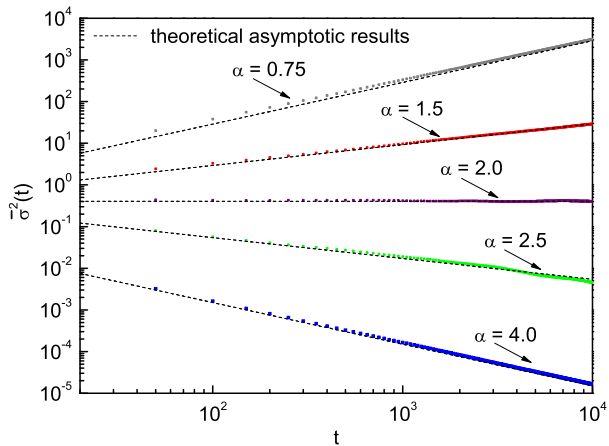}
\caption{(Color online) Asymptotic  long-time behavior of the variance $\bar{\sigma}^2(t)$ simulated for various values of the parameter $\alpha= U_0/(2 k_B T) + 1/2$, for the potential $U(x) = (U_0/2) \ln(1+x^2)$.  The dashed lines are the analytical predictions given by  Eq.~\eqref{D_13}. The simulation data is for $k_B T=1$, $\gamma=1$.}
\label{Fig_5_2}
\end{figure}

The explicit expression \eqref{D_13} for the time-averaged variance $\bar{\sigma}^2(t)$ gives important information about the ergodicity of the diffusion process in a logarithmic potential: the process is (mean)-ergodic if and only if $\bar{\sigma}^2(t)$ vanishes at long times  \cite{Pap91}. An analysis of Eq.~\eqref{D_13} reveals that  $\bar{\sigma}^2(t)\simeq t^X$ when $t\rightarrow \infty$, with $X=1$ for $\alpha<1$, $X=2-\alpha$ for $1<\alpha<3$, and $X =-1$ for $\alpha>3$ (see also  Figs.~\ref{Fig_5_1} and \ref{Fig_5_2}). Accordingly,  the process  is ergodic for $\alpha > 2$ ($k_B T / U_0 < 1/3$), while ergodicity is broken  for $\alpha < 2$ ($k_B T / U_0 > 1/3$). The ergodicity of the process for $\alpha > 2$ is in agreement with the Khinchin theorem \cite{Khi49}, since in this case there exists a stationary correlation function Eq.~\eqref{C_9} which vanishes as time tends to infinity. For $2 < \alpha < 3$ ($1/5 < k_B T / U_0 < 1/3$), even though the system is  ergodic, the slow decay of $\bar{\sigma}^2(t)$ means that ergodicity is reached anomalously slowly.
The nonergodic behavior for $\alpha < 1$ is not surprising, since in this regime, there exists no stationary equilibrium state. For $1 < \alpha < 2$, on the other hand, we observe broken ergodicity even though the Boltzmann equilibrium distribution is normalizable. This is due to the fact that the second moment increases with time \cite{Bar10} and thus diverges for the equilibrium distribution.
The asymptotic power-law exponent of $\bar{\sigma}^2(t)$ is shown in Fig.~\ref{Fig_5_1}. The analytic prediction \eqref{D_13} perfectly matches the numerical results, except near the points $\alpha = 1$ and $\alpha = 3$, where the convergence of the Langevin simulations is very slow and we expect transient logarithmic corrections. The asymptotic algebraic behavior of the time-averaged variance is further confirmed in the double-logarithmic plot presented in Fig.~\ref{Fig_5_2}. We again observe perfect agreement between analytics and numerics for different values of the parameter $\alpha$.

\begin{figure}
\includegraphics[trim=20mm 10mm 25mm 15mm, clip, width=0.47\textwidth]{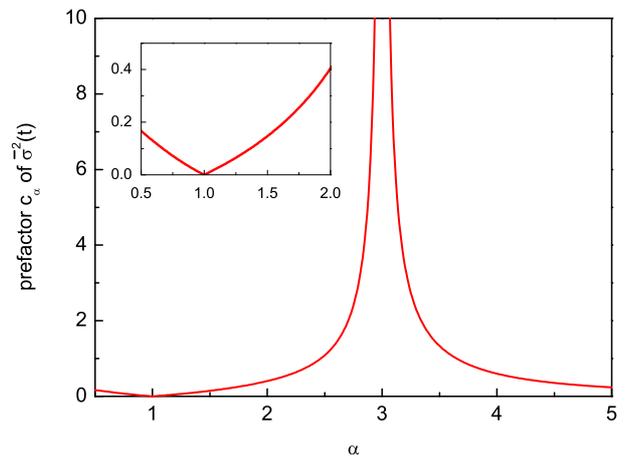}
\caption{(Color online) Prefactor $c_{\alpha}$ of $\bar{\sigma}^2(t) \simeq c_{\alpha} (4 D t)^{X}$, Eq.~\eqref{D_13}, as a function of $\alpha$ for $U(x) = (U_0/2) \ln(1+x^2)$. Note that it is zero at the transition from normal to subdiffusion ($\alpha = 1$) and diverges at $\alpha = 3$. Inset: Detail for $\alpha < 2$. The result for $\alpha >3$ was taken from \cite{Dec11}.}
\label{Fig_5_3}
\end{figure}

For $\alpha < 3$, the results in Eq.~\eqref{D_13} can be fully obtained from our asymptotic analysis and depend on the explicit form of the potential $U(x)$ only through the partition function $Z$ (for $1 < \alpha < 3$) or not at all (for $\alpha < 1$). For $\alpha > 3$, on the other hand, the contribution we get from the asymptotic behavior of the potential is of the same order as the one stemming from the behavior of the potential for $x \sim \mathcal{O}(1)$. The asymptotic analysis, while correctly predicting the $\bar{\sigma}^2(t) \propto t^{-1}$ behavior, thus fails in reproducing the prefactor $c_{\alpha}$. In terms of the asymptotic analysis, $c_{\alpha}$ appears to depend on the time scale $\Delta$ (see Appendix \ref{SEC_C}), which determines how long it takes for the asymptotic results to accurately describe the behavior of the system and thus can only be obtained by comparing the asymptotic results to the exact solution. However, $c_{\alpha}$ can be obtained from the Boltzmann equilibrium distribution $W_{\text{eq}}(x)$ (we give a general expression valid for $\alpha > 3$ and potentials that are more strongly binding than the logarithmic one in \cite{Dec11}), which by definition depends on the shape of the potential $U(x)$ in the whole space. Figure~\ref{Fig_5_3} shows the prefactor of $\bar{\sigma}^2(t)$ for the specific choice of the potential $U(x) = (U_0/2) \ln(1+x^2)$. This can be interpreted as a generalized diffusion coefficient for the time average: $\bar{\sigma}^2(t) \simeq c_{\alpha} (4 D t)^{X}$.

\section{Conclusion} \label{SEC_6}
We have performed a detailed analysis of Brownian motion in an asymptotically logarithmic potential. 
We have obtained explicit expressions for 
the position correlation function \eqref{C_6} and the variance of the time-averaged position \eqref{D_13}. The asymptotic time dependence of these quantities is determined by the single parameter $\alpha= U_0/(2 k_B T) + 1/2$, which measures the ratio of the potential depth $U_0$ and the temperature. Both diffusion and ergodic properties are controlled by $\alpha$, and the system exhibits a surprising variety of different behaviors. The ergodic and aging properties of the system are closely related to the occurrence of the non-normalizable infinite covariant density (ICD). We note that the dependence of the asymptotic dynamics on the potential depth $U_0$ is a peculiarity of the logarithmic potential, and is not obtained for general, e.g. power-law, potentials. 

For shallow potentials, $\alpha < 1$, diffusion is normal which implies $\langle x^2(t) \rangle \propto t$, a nonstationary aging position correlation and broken ergodicity $\bar{\sigma}^2(t) \propto t$. In this case, there is no stationary distribution, so the time-dependent solution determines all moments. For medium strength potentials, $1 < \alpha < 2$, the system exhibits subdiffusion \cite{Lev05}, $\langle x^2(t) \rangle \propto t^{2-\alpha}$, while the correlation function is still aging and ergodicity is broken, $\bar{\sigma}^2(t) \propto t^{2-\alpha}$. In this regime, the non-normalizable ICD determines the second moment of the position. The ICD is also essential for determining the correlation function, as the Boltzmann equilibrium density yields an infinite value for the latter. We find a super-aging correlation function $C(t,t_0)$, which behaves as $C(t,t_0) \simeq \langle x^2(t_0) \rangle f((t-t_0)/t_0)$ with a prefactor $\langle x^2(t_0) \rangle$ which increases with time. For deep potentials, $2 < \alpha < 3$, diffusion freezes, $\langle x^2(t) \rangle \propto \textit{constant}$, and the correlation function has a stationary limit $C(t,t_0) \simeq C(t-t_0)$, which can now be determined from the Boltzmann equilibrium distribution. In this regime, the process  is ergodic, although the decay of $\bar{\sigma}^2(t) \propto t^{2-\alpha}$ is slow. Only for very deep potentials, $\alpha > 3$ do we obtain $\bar{\sigma}^2(t) \propto t^{-1}$, which is the behavior expected from the usual (e.g. harmonic) confining potentials \cite{Dec11}.

{\bf Acknowledgements:} This work was supported by the Israel Science Foundation, the Emmy Noether Program
of the DFG (contract No LU1382/1-1), the cluster of excellence Nanosystems Initiative Munich and the Focus Area Nanoscale of the FU Berlin. We also thank the Alexander von Humboldt-Foundation for support.

\appendix

\section{Discussion of the center part of the probability density and correlation function} \label{SEC_A}

\begin{table*}
\begin{align}
\begin{array}{|l|c|c|c|c|}
\hline  & x>1, x_0>1 & x>1, x_0<1 & x<1, x_0>1 & x<1, x_0<1 \\ 
\hline \tau \gg t_0 & 1 & t_0^{-\frac{3}{2}} & \tau^{-\frac{3}{2}} & \tau^{-\frac{3}{2}} t_0^{-\frac{3}{2}} \\ 
\hline t_0 \gg \tau & 1 & \begin{array}{lr} \tau^{-\frac{3}{2}}, &\alpha > 2 \\ \tau^{\frac{1}{2}-\alpha} t_0^{\alpha-2}, & \alpha < 2 \end{array} & \begin{array}{lr} \tau^{-\frac{3}{2}}, &\alpha > 2 \\ \tau^{\frac{1}{2}-\alpha} t_0^{\alpha - 2}, & \alpha < 2 \end{array} & \begin{array}{lr} \tau^{-3}, &\alpha > 2 \\ \tau^{-1-\alpha} t_0^{\alpha-2}, & \alpha < 2 \end{array} \\ 
\hline
\end{array} \nonumber
\end{align}
\caption{The contributions from the different parts of the correlation function relative to the contribution from $x>1$, $x_0>1$}
\label{TAB_2}
\end{table*}

In the main body of the work, we only discussed the tail part, that is the case $x > 1$ and $x_0>1$, of the probability density $P(x,t \vert x_0,t_0)$, arguing that the contribution of the other cases ($x < 1$ or $x_0 < 1$, or both) to the correlation function is negligibly small. In order to determine the probability density for those cases, we have to use expression \eqref{A_16} instead of \eqref{A_12} for the wave functions including the respective variables in the expansion \eqref{B_2}. So for the case $x > 1$ and $x_0 < 1$, for example, we have for Eq.~\eqref{B_2},
\begin{align}
&P(x,t \vert x_0, t_0) \simeq A^2 x^{1-2\alpha} + 2 \frac{x^{1-\alpha}}{\tilde{\psi}_{0}(x_0)} \int_{0}^{\infty} dk \nonumber \\
&\quad \times \left[ A_k^2 \tilde{\psi}_{k,e}(x_0) \left(a_{1 k} J_{\alpha}(k x) + a_{2 k} J_{-\alpha}(k x)\right) e^{-D k^2 \tau} \right. \nonumber \\
&\quad \; + B_k^2 \tilde{\psi}_{k,o}(x_0)\left. \left(b_{1 k} J_{\alpha}(k x) + b_{2 k} J_{-\alpha}(k x)\right) e^{-D k^2 \tau} \right] \label{E_1} \ .
\end{align}
We now employ the same analysis as we have used for the tail part (see section \ref{SEC_3} and in particular Eq.~\eqref{B_4}) in order to determine the leading order for long times. We then use a small-$k$ expansion of the even and odd wave functions in the center, Eq.~\eqref{A_16},
\begin{align}
\tilde{\psi}_{k,e}(x_0) &\simeq A_k \ ,\label{E_1a} \quad \tilde{\psi}_{k,o}(x_0) \simeq B_k k x_0 \ .
\end{align}
Using Eq.~\eqref{E_1a} (with the coefficients given by Eqs.~\eqref{A_20}, \eqref{A_25a}, \eqref{A_25b} and \cite{Gra07}, Eq.~(6.631.1)), we obtain for $x>1$, $x_0<1$ to leading order in $\tau$, once again dividing into even and odd contribution,
\begin{align}
P_{e}&(x, t \vert x_0,t_0) \nonumber \\
&\simeq \left\lbrace
\begin{array}{ll}
\frac{1}{Z \, \Gamma(\alpha)} x^{1-2 \alpha} \Gamma \left( \alpha,\frac{x^2}{4 D \tau} \right) \; &\text{for} \; \alpha > 1 \\[2ex]
\frac{1}{\Gamma(1-\alpha)} (4 D \tau)^{\alpha - 1} x^{1-2\alpha} e^{-\frac{x^2}{4 D \tau}} \; &\text{for} \; \alpha < 1 \ ,
\end{array} \right.  \nonumber \\[2ex]
P_{o}&(x, t \vert x_0,t_0) \simeq \frac{2}{\Gamma(\alpha)} (4 D \tau)^{-1-\alpha} x x_0 e^{-\frac{x^2}{4 D \tau}} \ .
\label{E_3}
\end{align}
The other cases can be obtained in the same way and we find for $x<1$, $x_0>1$,
\begin{align}
P_{e}&(x, t \vert x_0,t_0) \nonumber \\
&\simeq \left\lbrace
\begin{array}{ll}
\frac{1}{Z \, \Gamma(\alpha)} \Gamma \left( \alpha,\frac{x_0^2}{4 D \tau} \right) \; &\text{for} \; \alpha > 1 \\[2ex]
\frac{1}{\Gamma(1-\alpha)} (4 D \tau)^{\alpha - 1} e^{-\frac{x_0^2}{4 D \tau}} \; &\text{for} \; \alpha < 1 \ ,
\end{array} \right.   \nonumber \\[2ex]
P_{o}&(x, t \vert x_0,t_0) \simeq \frac{2}{\Gamma(\alpha)} (4 D \tau)^{-1-\alpha} x x_0^{2 \alpha} e^{-\frac{x_0^2}{4 D \tau}}
\label{E_4}
\end{align}
and for $x<1$, $x_0<1$,
\begin{align}
P_{e}&(x, t \vert x_0,t_0) \nonumber \\
&\simeq \left\lbrace
\begin{array}{ll}
\frac{1}{Z} \; &\text{for} \; \alpha > 1 \\[2ex]
\frac{1}{\Gamma(1-\alpha)} (4 D \tau)^{\alpha - 1} \; &\text{for} \; \alpha < 1 \ , 
\end{array} \right. \nonumber \\[2ex]
P_{o}&(x, t \vert x_0,t_0) \simeq \frac{4 \alpha}{\Gamma(\alpha)} (4 D \tau)^{-1-\alpha} x x_0 \ . \label{E_5}
\end{align}
The even part of Eq.~\eqref{E_5} also yields the long time limit of the probability density $P(x,\tau)$ for $x<1$. Using the above expressions for the probability density, we can calculate the three remaining contributions to the correlation function Eq.~\eqref{C_1} for long times and $\alpha >1$,
\begin{align}
4 &\int_{1}^{\infty} dx \int_{0}^{1} dx_0 \, x x_0 P_{o}(x,t \vert x_0,t_0) W(x_0,t_0) \nonumber \\
&\simeq \frac{2\sqrt{\pi}}{3 Z \, \Gamma(\alpha)} (4 D (t-t_0))^{\frac{1}{2}-\alpha} \nonumber \ , \\[2ex]
4 &\int_{0}^{1} dx \int_{1}^{\infty} dx_0 \, x x_0 P_{o}(x,t \vert x_0,t_0) W(x_0,t_0) \nonumber \\
&\simeq \frac{8}{3 Z \, \Gamma^2(\alpha)} (4 D (t-t_0))^{\frac{1}{2}-\alpha} h_{\alpha} \left(\frac{t-t_0}{t_0}\right) \nonumber \ , \\[2ex]
4 &\int_{0}^{1} dx \int_{0}^{1} dx_0 \, x x_0 P_{o}(x,t \vert x_0,t_0) W(x_0,t_0) \nonumber \\
&\simeq \frac{8 \alpha}{9 Z \, \Gamma(\alpha)} (4 D (t-t_0))^{-1-\alpha} \ , \label{E_6}
\end{align}
with
\begin{align}
h_{\alpha}(s) = \int_{0}^{\infty} dy \, y^2 e^{-y^2} \Gamma(\alpha, y^2 s) \ , \label{E_7}
\end{align}
which is approximately constant for small $s$ and behaves as $s^{-3/2}$ for large $s$. For $\alpha<1$, we then have
\begin{align}
4 &\int_{1}^{\infty} dx \int_{0}^{1} dx_0 \, x x_0 P_{o}(x,t \vert x_0,t_0) W(x_0,t_0) \nonumber \\
&\simeq \frac{2 \sqrt{\pi}}{3 \Gamma(\alpha) \Gamma(1-\alpha)} (4 D (t-t_0))^{\frac{1}{2}-\alpha} (4 D t_0)^{\alpha-1} \nonumber \ , \\[2ex]
4 &\int_{0}^{1} dx \int_{1}^{\infty} dx_0 \, x x_0 P_{o}(x,t \vert x_0,t_0) W(x_0,t_0) \nonumber \\
&\simeq \frac{2 \sqrt{\pi}}{3 \Gamma(\alpha)\Gamma(1-\alpha)} (4 D (t-t_0))^{\frac{1}{2}-\alpha} (4 D t_0)^{\alpha-1}  \nonumber \\
& \qquad \qquad \qquad \times j_{\alpha} \left(\frac{t-t_0}{t_0}\right) \nonumber , \\[2ex]
4 &\int_{0}^{1} dx \int_{0}^{1} dx_0 \, x x_0 P_{o}(x,t \vert x_0,t_0) W(x_0,t_0) \nonumber \\
&\simeq \frac{8 \alpha}{9 \Gamma(\alpha) \Gamma(1-\alpha)} (4 D (t-t_0))^{-1-\alpha} (4 D t_0)^{\alpha-1} \ , \label{E_7a}
\end{align}
with $j_{\alpha}(s) = (1+s)^{-3/2}$.
In order to compare the above expressions to the contribution from the tail part of the probability densities Eqs.~\eqref{C_8} and \eqref{C_9}, we summarize the dependencies on $\tau = t-t_0$ and $t_0$ relative to the contribution from $x>1$, $x_0>1$ in Tab.~\eqref{TAB_2}. The contributions for $x<1$ or $x_0<1$ are negligible for long times $\tau$ and $t_0$ and Eq.~\eqref{C_6} indeed gives the leading order of the correlation function.

\section{Discussion of the functions $f_{\alpha}(s)$ and $g_{\alpha}(s)$} \label{SEC_D}
In Eq.~\eqref{C_7}, we introduced the two functions $f_{\alpha}(s)$ and $g_{\alpha}(s)$,
\begin{align}
f_{\alpha}(s) &= \int_{0}^{\infty} dy \, y^{2} e^{-y^2} \, {}_{1}\text{F}_{1} \left( \frac{3}{2}; \alpha + 1; y^2 \right) \Gamma \left( \alpha, y^2 s \right) \ , \nonumber \\
g_{\alpha}(s) &=  \int_{0}^{\infty} dy \, y^{2} e^{-y^2} \, {}_{1}\text{F}_{1} \left( \frac{3}{2}; \alpha + 1; y^2 \right) e^{-y^2 s} \ .\label{F_1}
\end{align}
We now want to find simpler expressions for these functions that are valid in the limit of small and large $s$, respectively, and thus give us the limiting behavior of the correlation function Eqs.~\eqref{C_8} and \eqref{C_9}. First, we divide the integral at $y=1$ into,
\begin{align}
&f_{\alpha}(s) = \mathfrak{I}_1(s) + \mathfrak{I}_2(s) \nonumber \\
&= \left[\int_{0}^{1} + \int_{1}^{\infty}\right] dy \, y^{2} e^{-y^2} \, {}_{1}\text{F}_{1} \left( \frac{3}{2}; \alpha + 1; y^2 \right) \Gamma \left( \alpha, y^2 s \right). \label{F_3}
\end{align}
For small $s$, the argument of the incomplete Gamma function in the first integral is small and we may approximate $\Gamma(\alpha, y^2 s) \simeq \Gamma(\alpha)$, so that,
\begin{align}
\mathfrak{I}_1(s) \simeq \int_{0}^{1} dy \, y^{2} e^{-y^2} \, {}_{1}\text{F}_{1} \left( \frac{3}{2}; \alpha + 1; y^2 \right) \Gamma(\alpha) \ , \label{F_4}
\end{align}
which is a constant independent of $s$. In the second integral, we introduce the variable $z = \sqrt{s} \, y$ to get,
\begin{align}
\mathfrak{I}_2(s) = s^{-\frac{3}{2}} \int_{\sqrt{s}}^{\infty} dz \, z^{2} e^{-\frac{z^2}{s}} \, {}_{1}\text{F}_{1} \left( \frac{3}{2}; \alpha + 1; \frac{z^2}{s}\right) \Gamma(\alpha, z^2) . \label{F_5}
\end{align}
Since the argument of the hypergeometric function is large except for very small $z$ (which gives us another constant contribution), we may use the large-argument expansion \cite{Abr72}, Eq.~(13.5.1)
\begin{align}
{}_{1}\text{F}_{1} \left( \frac{3}{2}; \alpha + 1; y^2 \right) \simeq \frac{2 \Gamma(\alpha+1)}{\sqrt{\pi}} y^{1-2\alpha} e^{y^2} \label{F_6}
\end{align}
to write
\begin{align}
\mathfrak{I}_2(s) \simeq \frac{2 \Gamma(\alpha+1)}{\sqrt{\pi}} s^{\alpha - 2} \int_{\sqrt{s}}^{\infty} dz \, z^{3-2\alpha} \Gamma(\alpha, z^2) \ . \label{F_7}
\end{align}
This integral can be calculated and yields in the limit of small $s$,
\begin{align}
\mathfrak{I}_2(s) \simeq \frac{\Gamma(\alpha+1)}{\sqrt{\pi}(2 - \alpha)} s^{\alpha - 2} \left( 1 - \Gamma(\alpha) s^{2-\alpha} \right) \ . \label{F_8}
\end{align}
The limiting behavior of this expression now depends on the value of $\alpha$: For $\alpha > 2$ the second term dominates for small $s$ and $\mathfrak{I}_2(s)$ is constant, while for $\alpha < 2$, the first term dominates and we have $\mathfrak{I}_2(s) \propto s^{\alpha - 2}$. For small $s$, we thus have,
\begin{align}
f_{\alpha}(s) \simeq \left\lbrace
\begin{array}{ll}
\frac{\sqrt{\pi} \, \Gamma(\alpha+1) \Gamma(\alpha) \Gamma(\alpha-2)}{4 \Gamma^2 \left(\alpha - \frac{1}{2}\right)} \quad &\text{for} \quad \alpha > 2 \\[2 ex]
\frac{\Gamma(\alpha+1)}{\sqrt{\pi}(2 - \alpha)} s^{\alpha - 2} \quad &\text{for} \quad \alpha < 2 \ .
\end{array} \right. \label{F_9}
\end{align}
The value of the constant for $\alpha > 2$ is obtained by evaluating $f_{\alpha}(0)$ using Mathematica.

For large $s$, we introduce the variable $z = \sqrt{s} \, y$ in $\mathfrak{I}_1(s)$,
\begin{align}
\mathfrak{I}_1(s) = s^{-\frac{3}{2}} \int_{0}^{\sqrt{s}} dz \, z^{2} e^{-\frac{z^2}{s}} \, {}_{1}\text{F}_{1} \left( \frac{3}{2}; \alpha + 1; \frac{z^2}{s} \right) \Gamma(\alpha, z^2). \label{F_10}
\end{align}
The argument of the exponential and hypergeometric functions now is small except for very large $z$, for which the incomplete Gamma function is exponentially small (see below). For small arguments, both the exponential and hypergeometric functions are approximately $1$, so we have,
\begin{align}
\mathfrak{I}_1(s) \simeq s^{-\frac{3}{2}} \int_{0}^{\sqrt{s}} dz \, z^{2} \Gamma(\alpha, z^2) \ , \label{F_11}
\end{align}
which for large $s$ reduces to
\begin{align}
\mathfrak{I}_1(s) \simeq \frac{\Gamma \left(\alpha + \frac{3}{2} \right)}{3} s^{-\frac{3}{2}} \ . \label{F_12}
\end{align}
In $\mathfrak{I}_2(s)$, we expand the incomplete Gamma function for large arguments (\cite{Abr72}, Eq.~(6.5.32))
\begin{align}
\Gamma(\alpha, y) \simeq y^{\alpha-1} e^{-y} \label{F_13}
\end{align}
and use Eq.~\eqref{F_6} to get,
\begin{align}
\mathfrak{I}_2(s) \simeq \frac{2 \Gamma(\alpha+1)}{\sqrt{\pi}} s^{\alpha-1} \int_{1}^{\infty} dy \, y e^{-y^2 s} \ . \label{F_15}
\end{align}
Evaluating the integral gives,
\begin{align}
\mathfrak{I}_2(s) \simeq \frac{ \Gamma(\alpha+1)}{\sqrt{\pi}} s^{\alpha-2} e^{-s} \ , \label{F_16}
\end{align}
which vanishes exponentially for large $s$. So, for large $s$, we have,
\begin{align}
f_{\alpha}(s) \simeq \frac{\Gamma \left(\alpha + \frac{3}{2} \right)}{3} s^{-\frac{3}{2}} \ . \label{F_17}
\end{align}
From Eqs.~\eqref{F_9} and \eqref{F_17}, we have for $f_{\alpha}(s)$,
\begin{align}
f_{\alpha}(s) \simeq \left\lbrace
\begin{array}{ll}
\frac{\sqrt{\pi} \, \Gamma(\alpha+1) \Gamma(\alpha) \Gamma(\alpha-2)}{4 \Gamma^2 \left(\alpha - \frac{1}{2}\right)} \quad &\text{for} \; s \ll 1 \; \text{and} \; \alpha > 2\\
\frac{\Gamma(\alpha+1)}{\sqrt{\pi}(2 - \alpha)} s^{\alpha - 2} \quad &\text{for} \; s \ll 1 \; \text{and} \; \alpha < 2 \\
\frac{\Gamma \left(\alpha + \frac{3}{2} \right)}{3} s^{-\frac{3}{2}} \quad &\text{for} \; s \gg 1 \ .
\end{array} \right. \label{F_18}
\end{align}
In a similar manner, we obtain,
\begin{align}
g_{\alpha}(s) \simeq \left\lbrace
\begin{array}{ll}
\frac{\Gamma(\alpha+1)\Gamma(2-\alpha)}{\sqrt{\pi}}s^{-1} \quad &\text{for} \; s \ll 1 \\
\frac{\sqrt{\pi}}{4} s^{- \frac{1}{2}-\alpha } \quad &\text{for} \; s \gg 1 \ .
\end{array} \right. \label{F_19}
\end{align}

\section{Discussion of the integral over the correlation function in Eq.~(\ref{cint})} \label{SEC_C}
Before evaluating the double time integral over the correlation function $C(t'',t')$ in Eq.~\eqref{cint}, we have to address the problem that there are  regions where $\tau = t'' - t'$ is not large and thus the approximate expression Eq.~\eqref{C_6} is not valid. In these regions, we make use of the fact that $C(t'',t') \leq \langle x(t')^2 \rangle$ for $t'' > t'$ to provide an upper bound on the value of the correlation function. We write,
\begin{align}
\int_{t^{*}}^{t} &dt'' \int_{t^{*}}^{t} dt' \, C(t'',t') \nonumber \\
&= 2 \left[ \int_{t^{*}+\Delta}^{t} dt'' \int_{t^{*}}^{t''-\Delta} dt' \, C(t'',t') \right. \nonumber \\
&\quad \left. + \int_{t^{*}}^{t} dt'' \int_{t''-\Delta}^{t''} dt' \, C(t'',t') \right. \nonumber \\
&\quad \left. - \int_{t^{*}}^{t^{*}+\Delta} dt'' \int_{t''-\Delta}^{t^{*}} dt' \, C(t'',t') \right] \ .\label{D_5}
\end{align}
The first of the three integrals on the right-hand side now satisfies $\tau = t'' - t' \geq \Delta$ with $\Delta$ chosen such  that the approximation \eqref{C_6} of the correlation function is accurate. The time scale $\Delta$ can in principle be obtained by comparing the result \eqref{C_6} with the exact correlation function (obtained e.g. numerically). However, the asymptotic behavior will turn out to be independent of $\Delta$ in most cases. In the remaining two integrals, we use the estimate $C(t'',t') \leq \langle x(t')^2 \rangle$ and the expression \eqref{B_7} for the second moment to obtain,
\begin{align}
&\left[\int_{t^{*}}^{t} dt'' \int_{t''-\Delta}^{t''} dt' - \int_{t^{*}}^{t^{*}+\Delta} dt'' \int_{t''-\Delta}^{t^{*}} dt' \right] \, C(t'',t')  \nonumber \\
&\leq \left\lbrace
\begin{array}{ll}
\frac{2 \alpha - 1}{3 \alpha - 6} t \Delta \; &\text{for} \; \alpha > 2 \\
\frac{1}{Z \Gamma(\alpha) (2-\alpha) (3-\alpha)} (4 D t)^{-\alpha + 2} t \Delta \; &\text{for} \; 1 < \alpha < 2 \\
\frac{1-\alpha}{2} (4 D t) t \Delta \; &\text{for} \; \alpha < 1
\end{array} \right. \label{D_6}
\end{align}
in the limit of large $t$ (specifically $t \gg t^{*}$ and $t \gg \Delta$).
Using the general form of the correlation function \eqref{C_6},
\begin{align}
C(t'',t') \simeq (t'' - t')^{\mu} \phi \left( \frac{t''-t'}{t'} \right)\ , \label{D_7}
\end{align}
 we can express the first integral in Eq.~\eqref{D_5} as, 
\begin{align}
&\int_{t^{*}}^{t} dt'' \int_{t^{*}}^{t''-\Delta} dt' \, C(t'',t') \nonumber \\ 
&\simeq \int_{t^{*}}^{t} dt'' \int_{t^{*}}^{t''-\Delta} dt' \, (t'' - t')^{\mu} \phi \left( \frac{t''-t'}{t'} \right) \ .\label{D_8}
\end{align}
Making the change of  variable $s=(t''-t')/t'$ in the $t'$-integral, we further have
\begin{align}
\int_{t^{*}}^{t} dt'' t''^{\mu+1} \int_{\frac{\Delta}{t''-\Delta}}^{\frac{t''}{t^{*}}-1} ds \, \frac{s^{\mu}}{(s+1)^{\mu+2}} \phi(s) \ . \label{D_9}
\end{align}
where the function $\phi(s)$ is given by Eq.~\eqref{C_7},
\begin{align}
\phi(s) = \left\lbrace 
\begin{array}{ll} 
\int_{0}^{\infty} dy y^{2} e^{-y^2} {}_{1}\text{F}_{1} \left( \frac{3}{2}; \alpha + 1; y^2 \right) \\
\qquad \qquad \times \Gamma \left( \alpha, y^2 s \right) \; &\text{for} \; \alpha > 1 \\[2ex]
s^{1-\alpha}\int_{0}^{\infty} dy y^{2} {}_{1}\text{F}_{1} \left( \frac{3}{2}; \alpha + 1; y^2 \right) \\
\qquad \qquad \times e^{-y^2 (s+1)}  \; &\text{for} \; \alpha < 1 \ ,
\end{array} \right. \label{E_27} 
\end{align}
with the asymptotic behavior (see Appendix \ref{SEC_D}),
\begin{align}
\phi(s) \simeq \left\lbrace 
\begin{array}{ll} 
s^{-\frac{3}{2}} \; &\text{for} \; s \gg 1 \; \text{and} \; \alpha > 1 \\
s^{ - \frac{1}{2}-\alpha} \; &\text{for} \; s \gg 1 \; \text{and} \; \alpha < 1 \\
\text{const.} \; &\text{for} \; s \ll 1 \; \text{and} \; \alpha > 2 \\
s^{\alpha - 2} \; &\text{for} \; s \ll 1 \; \text{and} \; 1 <\alpha < 2 \\
s^{-1} \; &\text{for} \; s \ll 1 \; \text{and} \; \alpha < 1  \ .
\end{array} \right. \label{E_28} 
\end{align}
Note that for now we omit any $s$-independent prefactors of $\phi(s)$, these have been included in the result in Sec \ref{SEC_5}. The exponent $\mu$ is given by,
\begin{align}
\mu = \left\lbrace 
\begin{array}{ll} 
2-\alpha \; &\text{for} \; \alpha > 1 \\
1 \; &\text{for} \; \alpha < 1 \ .
\end{array} \right. \label{E_29} 
\end{align}
When $t''$ is close to the lower boundary, the $s$-integral vanishes since the lower and upper boundary are the same. For this reason the main contribution comes from $t''$ close to the upper boundary, that is $t'' \approx t$. Using $t \gg \Delta$ and $t \gg t^*$, we have for the $s$-integral
\begin{align}
\int_{\frac{\Delta}{t''}}^{\frac{t''}{t_0}} ds \, \frac{s^{\mu}}{(s+1)^{\mu+2}} \phi(s) \ . \label{E_30}
\end{align}
We split this integral at $s=1$ into,
\begin{align}
\int_{\frac{\Delta}{t''}}^{1} ds \, \frac{s^{\mu}}{(s+1)^{\mu+2}} \phi(s) + \int_{1}^{\frac{t''}{t_0}} ds \, \frac{s^{\mu}}{(s+1)^{\mu+2}} \phi(s) \ . \label{E_31}
\end{align}
In the first integral, we use the small argument expansion of $\phi(s)$, Eq.~\eqref{E_28} and obtain,
\begin{align}
\int_{\frac{\Delta}{t''}}^{1} &ds \, \frac{s^{\mu}}{(s+1)^{\mu+2}} \phi(s) \nonumber \\
& \simeq \left\lbrace 
\begin{array}{ll}
1 - \left( \frac{\Delta}{t''} \right)^{3-\alpha} \; &\text{for} \; \alpha > 2 \\[2 ex]
1- \frac{\Delta}{t''} \; &\text{for} \; \alpha < 2 \ .
\end{array} \right.\label{E_32}
\end{align}
For large $t''$, the leading contribution for the case $\alpha > 2$ depends on $\alpha$:
\begin{align}
\int_{\frac{\Delta}{t''}}^{1} &ds \, \frac{s^{\mu}}{(s+1)^{\mu+2}} \phi(s) \nonumber \\
& \simeq \left\lbrace 
\begin{array}{ll}
\left( \frac{\Delta}{t''} \right)^{3-\alpha} \; &\text{for} \; \alpha > 3 \\[2 ex]
\text{const.} \; &\text{for} \; 2 < \alpha < 3 \\[2 ex]
\text{const.} \; &\text{for} \; \alpha < 2 \ .
\end{array} \right.\label{E_33}
\end{align}
In the second integral in Eq.~\eqref{E_31}, we use the large argument expansion of $\phi(s)$,
\begin{align}
\int_{1}^{\frac{t''}{t_0}} &ds \, \frac{s^{\mu}}{(s+1)^{\mu+2}} \phi(s) \nonumber \\
& \simeq \left\lbrace 
\begin{array}{ll}
1 - \left( \frac{t''}{t_0} \right)^{-\frac{9}{2}} \; &\text{for} \; \alpha > 1 \\[2 ex]
1 - \left( \frac{t''}{t_0} \right)^{-\frac{7}{2}-\alpha} \; &\text{for} \; \alpha < 1 \ .
\end{array} \right.\label{E_34}
\end{align}
So, for large $t''$, we have
\begin{align}
\int_{1}^{\frac{t''}{t_0}} &ds \, \frac{s^{\mu}}{(s+1)^{\mu+2}} \phi(s) \simeq \text{const.} \ . \label{E_35}
\end{align}
In the limit $t \gg t_0$, we then get for the expression Eq.~\eqref{D_8},
\begin{align}
&\int_{t_0+\Delta}^{t} dt'' t''^{\mu+1} \int_{\frac{\Delta}{t''-\Delta}}^{\frac{t''}{t_0}-1} ds \, \frac{s^{\mu}}{(s+1)^{\mu+2}} \phi(s) \nonumber \\
&\simeq \left\lbrace \begin{array}{ll}
t^{1} \; &\text{for} \; \alpha > 3 \\[2 ex]
t^{4-\alpha} \; &\text{for} \; 1 < \alpha < 3 \\[2 ex]
t^3 \; &\text{for} \; \alpha < 1 \ .
\end{array} \right.  \label{E_36}
\end{align}
In order to obtain the prefactor to these asymptotic forms, we take the limit $t'' \rightarrow \infty$ in the $s$-integral in Eq.~\eqref{E_36} for $\alpha < 3$, since the integral is constant in this limit:
\begin{align}
\langle \bar{x}^2(t) \rangle &\simeq \frac{1}{t^2} \int_{0}^{t} dt'' t''^{\mu+1} \int_{0}^{\infty} ds \, \frac{s^{\mu}}{(s+1)^{\mu+2}} \phi(s) \nonumber \\
&\simeq \frac{1}{\mu + 2} \, t^{\mu} \int_{0}^{\infty} ds \, \frac{s^{\mu}}{(s+1)^{\mu+2}} \phi(s) \label{E_37}
\end{align}
For $\alpha > 3$, the $s$integral grows as $t''^3$ as the lower boundary approaches zero [see Eq.~\eqref{E_33}] and we may not take the lower boundary to zero. However, since the main contribution stems from small values of $s$, we may approximate $\phi(s) \approx \phi(0)$ and have,
\begin{align}
\langle \bar{x}^2(t) \rangle &\simeq \frac{1}{t^2} \, \phi(0) \int_{0}^{t} dt'' t''^{\mu+1} \int_{\frac{\Delta}{t''}}^{\infty} ds \, \frac{s^{\mu}}{(s+1)^{\mu+2}} \nonumber \\
&\simeq \frac{\phi(0)}{-\mu - 1} \, \frac{1}{t^2} \int_{0}^{t} dt'' t''^{\mu+1} \left( \frac{t''}{\Delta} \right)^{-\mu - 1} \nonumber \\
&\simeq \frac{\phi(0) \Delta^{\mu+1}}{-\mu - 1} \, t^{-1} \ . \label{E_38}
\end{align}
Summarizing these results, we have
\begin{align}
&\langle \bar{x}^2(t) \rangle_l \nonumber \\
&\simeq \left\lbrace 
\begin{array}{ll} 
\frac{f(0)}{\alpha - 3} \, (4 D \Delta)^{2-\alpha} \left( \frac{t}{\Delta} \right)^{-1} \\
\qquad \qquad \; \text{for} \; \alpha > 3 \\[2 ex]
\frac{\sqrt{\pi}}{Z \, \Gamma(\alpha+1) \Gamma(\alpha)(4 - \alpha)} \, (4 D t)^{2-\alpha} \int_{0}^{\infty} ds \, \frac{s^{2-\alpha}}{(s+1)^{4-\alpha}} \, f_{\alpha}(s) \\
\qquad \qquad \; \text{for} \; 1 < \alpha < 3 \\[2 ex]
\frac{\sqrt{\pi}}{3 \Gamma(\alpha) \Gamma(1-\alpha)} \, 4 D t \int_{0}^{\infty} ds \, \frac{s}{(s+1)^{3}} \, g_{\alpha}(s) \\
\qquad \qquad \; \text{for} \; \alpha < 1 \ ,
\end{array} \right. \label{D_10}
\end{align}
where we use the subscript $l$ to denote the contribution from the long-time behavior of the correlation function. The functions $f_{\alpha}(s)$ and $g_{\alpha}(s)$ are the ones defined in Eq.~\eqref{C_7}. The other contribution comes from the short-time correlation function, which we have approximated by a constant in Eq.~\eqref{D_6},
\begin{align}
\langle \bar{x}^2(t) \rangle_s
\leq \left\lbrace 
\begin{array}{ll}
\frac{2 \alpha - 1}{3 \alpha - 6} \left( \frac{t}{\Delta} \right)^{-1} \; &\text{for} \; \alpha > 2 \\[2ex]
\frac{1}{Z \, \Gamma(\alpha)(2-\alpha) (3-\alpha)} \\
\qquad \times (4 D \Delta) (4 D t)^{1-\alpha} \; &\text{for} \; 1 < \alpha < 2 \\[2ex]
\frac{1-\alpha}{2} (4 D \Delta) \; &\text{for} \; \alpha < 1 \ .
\end{array} \right. \label{D_11}
\end{align}
Comparing the long and short time contributions, Eqs.~\eqref{D_10} and \eqref{D_11}, we see that the long-time contribution dominates for $\alpha < 3$, while both contributions are of the same order for $\alpha > 3$. As a consequence, we have,
\begin{align}
&\langle \bar{x}^2(t) \rangle \nonumber \\
&\simeq \left\lbrace 
\begin{array}{ll}
c_{\alpha} (4 D t)^{-1} \\
\qquad \qquad \; \text{for} \; \alpha > 3 \\[2ex]
\frac{\sqrt{\pi}}{Z \, \Gamma(\alpha+1) \Gamma(\alpha)(4 - \alpha)} \, (4 D t)^{2-\alpha} \int_{0}^{\infty} ds \, \frac{s^{2-\alpha}}{(s+1)^{4-\alpha}} \, f_{\alpha}(s) \\
\qquad \qquad \; \text{for} \; 1 < \alpha < 3 \\[2 ex]
\frac{\sqrt{\pi}}{3 \Gamma(\alpha) \Gamma(1-\alpha)} \, 4 D t \int_{0}^{\infty} ds \, \frac{s}{(s+1)^{3}} \, g_{\alpha}(s) \\
\qquad \qquad \; \text{for} \; \alpha < 1 \ .
\end{array} \right.
\end{align}
The prefactor $c_{\alpha}$ for $\alpha > 3$ cannot be obtained within this asymptotic analysis, since it will depend in general on the shape of the potential $U(x)$ in the whole space instead of just the logarithmic $U(x) \simeq U_0 \ln (x)$ asymptotic large $x$ behavior. $c_\alpha$ can, however, be obtained from the Boltzmann equilibrium distribution \cite{Dec11}.

\bibliography{BIB_APS-2}

\end{document}